%% file: main.tex
\documentclass[sigconf]{acmart}

\usepackage{fancyhdr}
\usepackage{algorithmic}
\usepackage{graphicx}
\usepackage{textcomp}
\usepackage{xcolor}
\usepackage{bm}
\usepackage{booktabs}
\usepackage[ruled,linesnumbered]{algorithm2e}
\usepackage{multirow}
\usepackage{enumitem}
\usepackage{microtype}
\usepackage{graphicx}
\usepackage{subfigure}
\usepackage{booktabs} 
\usepackage{amsfonts} 

\usepackage{amssymb}
\usepackage{amsmath}
\usepackage{sansmath}
\usepackage{subfigure}
\usepackage{amsthm}
\usepackage{tcolorbox}
\usepackage{colortbl}
\usepackage{tikz}
\usepackage{url}
\usepackage{pifont}
\usepackage{graphicx}
\usepackage{makecell}

\usepackage{pgfplots}

\usepackage{cleveref}

\usepackage{amsmath}

\newcommand{\floor}[1]{\left\lfloor #1\right\rfloor}

\crefformat{section}{\S#2#1#3} 
\crefformat{subsection}{\S#2#1#3}
\long\def\comment#1{}

\definecolor{LightCyan}{rgb}{0.88,1,1}
\definecolor{LightGreen}{HTML}{DBEDC5}
\definecolor{LightRed}{rgb}{1,0.88,1}
\definecolor{LightYellow}{rgb}{1,1,0.88}
\definecolor{LightGray}{gray}{0.9}

\input{command}

\AtBeginDocument{%
  }

\setcopyright{acmlicensed}
\copyrightyear{2018}
\acmYear{2018}
\acmDOI{XXXXXXX.XXXXXXX}

\acmConference[Conference acronym 'XX]{Make sure to enter the correct
  conference title from your rights confirmation email}{June 03--05,
  2018}{Woodstock, NY}
\acmISBN{978-1-4503-XXXX-X/18/06}

\begin{document}

\title{CardOOD: Robust Query-driven Cardinality Estimation under Out-of-Distribution}

\author{Rui Li}
\orcid{0000-0001-5109-3700}
\affiliation{%
  \institution{The Chinese University of Hong Kong}
  \city{Hong Kong}
  \country{China}
}
\email{lirui@se.cuhk.edu.hk}

\author{Kangfei Zhao}
\orcid{0000-0001-5109-3700}
\affiliation{%
  \institution{Beijing Institute of Technology}
  \city{Beijing}
  \country{China}
}
\email{zkf1105@gmail.com}

\author{Jeffrey Xu Yu}
\orcid{0000-0001-5109-3700}
\affiliation{%
  \institution{The Chinese University of Hong Kong}
  \city{Hong Kong}
  \country{China}
}
\email{yu@se.cuhk.edu.hk}

\author{Guoren Wang}
\orcid{0000-0001-5109-3700}
\affiliation{%
  \institution{Beijing Institute of Technology}
  \city{Beijing}
  \country{China}
}
\email{wanggrbit@126.com}



\begin{abstract}
Query-driven learned estimators are accurate, flexible, and lightweight alternatives to traditional estimators in query optimization. 
However, existing query-driven approaches struggle with the Out-of-distribution (OOD) problem, where the test workload distribution differs from the training workload, leading to performance degradation.
In this paper, we present \CardOOD, a general learning framework designed to construct robust query-driven cardinality estimators that are resilient against the OOD problem. Our framework focuses on offline training algorithms that develop one-off models from a static workload, suitable for model initialization and periodic retraining. In \CardOOD, we extend classical transfer/robust learning  techniques to train query-driven  cardinality estimators, and the algorithms fall into three categories: representation learning, data manipulation, and new learning strategies. 
As these learning techniques are originally evaluated in computer vision tasks, we also propose a new learning algorithm that exploits the property of cardinality estimation.  
%
This algorithm, lying in the category of new learning strategy, models the partial order constraint of cardinalities by a self-supervised learning task.  
Comprehensive experimental studies demonstrate the efficacy of the algorithms of \CardOOD in mitigating the OOD problem to varying extents. We further integrate \CardOOD into \PostgreSQL, showcasing its practical utility in query optimization. 
\end{abstract}



\maketitle

\section{Introduction}
Cardinality estimation is a fundamental task in query optimization. In the early years, RDBMSs adopted statistical methods~\cite{DBLP:conf/vldb/PoosalaI97, DBLP:journals/vldb/GunopulosKTD05, DBLP:conf/sigmod/PoosalaIHS96} to estimate the cardinalities of SQL queries, where impractical statistical assumptions, i.e., uniform and independence of attributes, lead to inaccurate estimation results and suboptimal query plans~\cite{DBLP:conf/sigmod/IoannidisC91}. Recent studies employ learning approaches for cardinality estimation, which are shifting from the Machine Learning (ML) approaches~\cite{DBLP:conf/sigmod/MaT19, DBLP:journals/pvldb/KieferHBM17, DBLP:journals/pvldb/DuttWNKNC19} to Deep Learning (DL) approaches~\cite{DBLP:journals/pvldb/SunL19, DBLP:conf/cidr/KipfKRLBK19,  DBLP:journals/pvldb/LiuD0Z21, DBLP:conf/icde/Thirumuruganathan20, DBLP:journals/pvldb/YangLKWDCAHKS19, 
  DBLP:conf/sigmod/HasanTAK020, DBLP:journals/pvldb/LuKKC21, DBLP:journals/pvldb/ZhaoCSM22}.
The shifting is driven by the powerful approximation capability of neural networks in end-to-end applications and high-end DL frameworks. 
All the DL approaches can improve the estimation accuracy as well as the quality of generated query plans. 

Existing DL-based cardinality estimation approaches are broadly categorized into data-driven and query-driven approaches.
Data-driven approaches learn the joint probability distribution of all attributes by deep density models, i.e., deep autoregressive neural networks~\cite{DBLP:conf/icde/Thirumuruganathan20, DBLP:journals/pvldb/YangLKWDCAHKS19,
  DBLP:conf/sigmod/HasanTAK020}, Sum-Product Network~\cite{DBLP:journals/pvldb/HilprechtSKMKB20}, and cardinality estimation is implemented by local probability density estimation. It is difficult to deploy these learned estimators in DBMSs because of the huge training cost and inference inefficiency. Furthermore, these approaches typically do not support complex DB schema and queries such as cyclic joins or self-joins~\cite{DBLP:journals/pvldb/NegiWKTMMKA23}.

Another line of DL-based cardinality estimation is the query-driven approaches that build a regression model which maps features of a query to its corresponding cardinality in a training workload.
In contrast to data-driven approaches, the training and inference of query-driven models are time and space-efficient due to low model complexity. And they are flexible to support  complex SQL queries, by transforming the query into hidden vector representation by various neural network architectures. To the best of our knowledge, query-driven learned estimators are more practical to deploy in real DBMSs~\cite{DBLP:journals/pvldb/NegiWKTMMKA23}. 
However,  existing learned estimator are trained by optimizing the empirical error of the training workload under the assumption that the training and test workload are i.i.d.  
The largest challenge that query-driven approaches face is they are sensitive to workload shift, i.e., the distribution of test query is shifting from that of the training workload.

Recent studies~\cite{DBLP:conf/sigmod/LiLK22,  DBLP:journals/pacmmod/WuI24} have proposed methods of statistical hypothesis testing to
detect  distribution shift of the workloads. 
Out-of-distribution (OOD) test queries are detected regarding the distribution shift of query features in the
workload, lying in the literals in the filter predicates, the combinations of filter predicates/join conditions 
and the combinations of base tables, etc. 
For instance, the commonly used benchmark DSB~\cite{DBLP:conf/dsb/ding2021} includes several hand-written query templates with different
filtering predicates and join graphs. 
The test workload can be regarded as an OOD workload if it contains different query templates with the training workload. 
The OOD problem derives from the lack of a sufficiently large, diverse and balanced training query set, and the query join graphs and/or the predicates may change over time in a database. 
Since models trained by previous query logs are difficult to generalize well on the OOD test queries, these studies~\cite{DBLP:conf/sigmod/LiLK22,  DBLP:journals/pacmmod/WuI24} have developed incremental learning-based approaches to maintain the learned estimators, balancing the cost of model retraining and performance degeneration. 

In this paper, our goal is to construct a general learning framework that can build robust query-driven cardinality estimators against the OOD problem.
%
%
Distinguished from the online learning approaches that digest a dynamically incremental workload~\cite{DBLP:conf/sigmod/LiLK22,  DBLP:journals/pacmmod/WuI24}, we concentrate on offline training algorithms that build one-off models from a stationary workload, which can be used for better model initialization and/or periodic retraining.  
In addition, these learning algorithms are orthogonal to the DL models for cardinality estimation. In other words, these learning algorithms can be applied to train any DL-based query-driven estimators.

%
%

%

We first extend typical transfer/robust learning techniques, beyond optimizing the empirical loss, to train cardinality estimators. 
These algorithms are categorized into three classes, i.e., representation learning, data manipulation, and new learning strategies.
More specifically, representation learning approaches tend to align the hidden representations among sub-distributions in the training query workload. Data manipulation approaches utilize data augmentation to enlarge and enrich the training queries. The learning strategy-based approaches use new learning strategies, e.g., Distributionally Robust Optimization~\cite{DBLP:journals/corr/abs-1911-08731}, self-supervised learning~\cite{DBLP:journals/pami/JingT21}.  
Although these learning techniques for OOD generalization have been evaluated in computer vision tasks, i.e., image classification, 
whether these techniques are also effective in building robust cardinality estimators remains unrevealed. 
The answer to this question is not straightforward.
In image classification, the features of an image commonly drift in terms of color, style, and lightness etc., while the principal visual features of the object in an image are invariant to be exploited during learning.
On the contrary, in cardinality estimation, query features drift anywhere, e.g., the predicates, the join condition, and the relations, which result in changes in the cardinality. 
Moreover, as a regression task, the prediction of query cardinality is more sensitive than the target image class, where any minor changes in query features may lead to changes in the cardinality of several magnitudes.

To further exploit the property of cardinality estimation, we propose a new learning algorithm based on self-supervised learning, in the category of learning strategy. In light of partial order embedding~\cite{DBLP:conf/icml/McFeeL09}, the algorithm models the partial order of queries regarding their cardinality by optimizing a self-supervision learning objective. This partial order constraint, which naturally exists in SQL query workloads, promotes the query embeddings subject to the same partial order that implicitly copes OOD of query features.  
We develop an extensible learning framework called \CardOOD that integrates all the learning algorithms we explore. 
In our experimental studies, these algorithms alleviate the OOD problem in cardinality estimation by different degrees.
Finally, we deploy \CardOOD in a real DBMS, \PostgreSQL, for query optimization.
The contributions of this paper are summarized as follows:


%

\begin{itemize}[leftmargin=*]
\item We extend classical transfer/robust learning algorithms to build robust DL-based cardinality estimators for SQL select-project-join (SPJ) queries. Here, some algorithms are originally domain adaption algorithms where features of the target domain are known while we assume the target test query workload is unknown. 
\item We propose a novel self-supervised learning algorithm, called OrderEmb, which models the partial order of queries regarding the cardinality, for training robust cardinality estimators.
\item We build a DL framework, \CardOOD, which integrates 6 robust learning algorithms and 2 types of DL models. We evaluate the characteristics of \CardOOD in multiple facets. 
\item We conduct a comprehensive experimental evaluation for \CardOOD on two neural estimators for single relation range queries and multi-relation join queries. 
\item We deploy \CardOOD in \PostgreSQL for query optimization.  Test results on \imdb, \dsb and \job queries show that models constructed by \CardOOD achieve up to $5.6\%$, $36.6\%$ and $4.6\%$ improvement in query execution time.
\end{itemize}

\stitle{Roadmap}: The following of this paper is organized as follows. \cref{sec:ps} gives our problem statement. In \cref{sec:method}, we introduce 6 ML algorithms for establishing robust data-driven cardinality estimators, which fall into 3 categories. In the following, we elaborate on the implementation and characteristics of our \CardOOD framework in \cref{sec:cardood}. \cref{sec:exp} reports the experimental results. We review the related works in \cref{sec:rw} and conclude the paper in \cref{sec:conclusion}.

\section{Problem Statement}
\label{sec:ps}

A relational database consists of a set of relations, $\{ R_1, R_2,
\cdots R_N\}$, where a relation $R_i$ has $d_i$ attributes such as
$R_i = (A_{1}, \cdots A_{d_i})$. Here, an attribute $A_j$ is either a
numerical attribute in a domain with a given range $[min_j, max_j]$ or a
categorical attribute with a discrete finite domain $\{c_1, c_2,
\cdots c_{m_j}\}$.

We study cardinality estimation for the select-project-join (SPJ) SQL
queries with conjunctive conditions. 
A selection on an attribute is either a range filter (i.e., $[lb_j,
  ub_j]$, denoting the condition $lb_j \leq A_{j} \leq ub_j$), if the
attribute is numerical, or \IN filter (i.e., $A_j~\IN~C $, $C \subset
\{ c_1, c_2, \cdots c_{m_j}\}$, denoting the condition $\exists c_k \in C,
A_j = c_k$), if the attribute is categorical.
A projection can be on any attribute and a join we support is 
equi-join. 
The primary-foreign key join (PK/FK join), and foreign-foreign key join (FK/FK join), are treated as special equi-join with extra
constraints. 
\comment{
An example of relational algebra is given below:
\begin{align}
\sigma_{(100 \leq R_1.A_1 \leq 200) \land (R2.A_2~\IN~\{ 15, 20, 25
  \})} \displaystyle{ (R_1 \mathop{\Join}_{\substack{R_1.A_3
      = R_2.A_3}} R_2)} \nonumber   
\end{align}
where $\sigma$ is the select operator and $\Join$ is a join operator.
}
\comment{
It is worth mentioning that we support general select-project-join SQL queries.
All the existing learned estimators do not support joins other than PK/FK joins.  
For example, \cite{DBLP:journals/corr/abs-2006-08109} does not support cyclic join queries and selection conditions on join attributes, and \cite{DBLP:journals/pvldb/HilprechtSKMKB20} does not support multiple selection conditions on join attributes. 
}

The cardinality of an SQL query $q$ is the number of resulting tuples,
denoted as $c(q)$. To learn a cardinality estimator, like 
existing works~\cite{DBLP:conf/cidr/KipfKRLBK19}, we require a set of pre-defined joinable attribute pairs $\{
(R_1.A_i, R_2.A_j),$ $\cdots \}$, in addition to a set of relations
$\{ R_1,\cdots R_N\}$.  For example, two relations $\{R_1, R_2\}$ with
joinable attribute pairs $\{(R_1.A_3, R_2.A_3)\}$.

\stitle{Query-driven Learned Cardinality Estimation}: Given a set of relations and a set of
joinable attribute pairs, learn a model $\mathcal{M}: q \mapsto
\mathbb{R}$ from a training query workload  $\Dataset =\{ (q_1, c(q_1)),$ $(q_2,
c(q_2)), \cdots\}$, where $q_i$ is an SQL SPJ  query
and $c(q_i)$ is its actual cardinality, to predict the cardinality for unseen queries.
We use \Qerror to evaluate the accuracy of the
estimated value: 
\begin{align}
\Qerror(q) = \max\bigg\{\frac{c(q)}{\hat{c}(q)}, \frac{\hat{c}(q)}{{c}(q)} \bigg\} 
\end{align}
Intuitively, \Qerror quantifies the factor by which the estimated
count ($\hat{c}(q_i)$) differs from the true count ($c(q_i)$).  It is
symmetrical and relative so that it provides statistical stability
for true counts of various magnitudes. Here, we assume $c(q) \geq 1$
and $\hat{c}(q) \geq 1$.

Given a training query workload $\Dataset$, existing query-driven cardinality estimators are regression models, which are trained by minimizing the mean-squared-error (MSE) loss over $\Dataset$ as shown in Eq.~\eqref{eq:loss:erm}, where $c_{\Theta}(q)$ is the predicted cardinality of a query $q$ with the model parameter $\Theta$.
\begin{align}
\mse(\Dataset; \Theta) &= \frac{1}{|\mathcal{D}|} \sum_{q \in \mathcal{D}} \norm{\log c(q) - \log c_{\Theta}(q)}^2 \label{eq:loss:erm} \\
&= \frac{1}{|\mathcal{D}|} \log^2 \frac{c(q)}{c_{\Theta}(q)} \nonumber
\end{align}
To achieve an average low relative error, the true cardinality $c(q)$ and predicted cardinality $c_{\Theta}(q)$ are transformed into a logarithmic scale.
%
%

Models are trained by mini-batch stochastic gradient descent (SGD), and the learning paradigm is known as \emph{Empirical Risk Minimization} (ERM)~\cite{DBLP:conf/nips/Vapnik91}.
Suppose the training queries are instance samples from a distribution 
$\Prob$, e.g., a query log, ERM approximates the expected loss in Eq.~\eqref{eq:loss:exp} over $\Prob$ by optimizing the empirical loss over $\Dataset$ in Eq.~\eqref{eq:loss:erm}.
\begin{align}
    \loss_{exp}(\Prob; \Theta) = \mathbb{E}_{q \sim \Prob} [\norm{\log c(q) - \log c_{\Theta}(q)}^2] \label{eq:loss:exp}
\end{align}
The learned models are expected to generalize to an unseen test query set $\Dataset^*$, under the assumption that $\Dataset^*$ and
$\Dataset$ are subject to the same distribution, i.e., $\Dataset^* \sim \Prob$.
This assumption ensures that optimizing the empirical loss on $\Dataset$ will consistently achieve a low risk $\mse(\Dataset^*; \Theta)$ on
the test queries.

However, in practice, the same distribution assumption is too ideal to hold. Queries in real databases are diversified regarding 
the \textit{query features} including the literal distribution in the filter predicates, the combinations of filter
predicate/join conditions and the combinations of base tables, etc. 
And the query workload may change over time, leading to a distribution shift on the query features. 
In this paper, we delve into learning paradigms beyond ERM, to establish learned estimators that perform robustly, when the distribution of test query workload deviates from that of the training query  workload.

\comment{
\begin{figure}[t]
 \centering
	\begin{tabular}[h]{c}
        \subfigure[Training Query Workload] {
				\includegraphics[width=0.43\columnwidth]{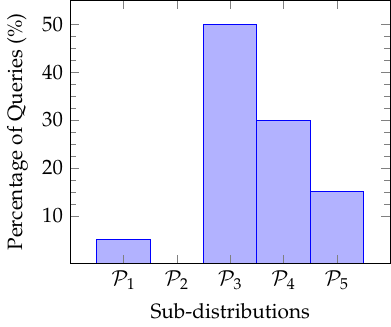}
			\label{fig:histogram:train}
		} 
        \subfigure[Test Query Workload] {
			\includegraphics[width=0.43\columnwidth]{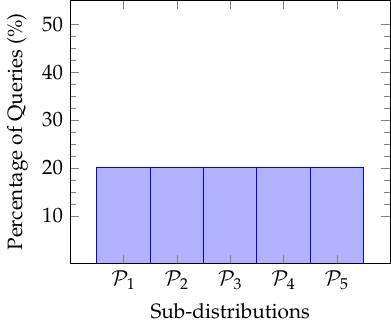}
			\label{fig:histogram:test}
		}
	\end{tabular}
	\caption{Query Distribution Shift Between Training and Test Query Workloads}
        \label{fig:distribution_shift}
\end{figure}
}

\stitle{Robust Learning Paradigms for Cardinality Estimation.} Given a training query workload $\Dataset \sim \Prob$, learn a model $\Model$ from $\Dataset$ that aims to achieve low test loss $\mse(\Dataset^*; \Theta)$ on a test query workload $\Dataset^* \sim \Prob^*$, where $\Prob^* \neq \Prob$. The test query workload $\Dataset^*$ and its underlying distribution $\Prob^*$ are unknown in the training. 
\comment{
Fig.~\ref{fig:distribution_shift} illustrates a toy example of the distribution shift of query workload. 
Here, $\{\Prob^*_{1}, \cdots, \Prob^*_{5}\}$ are five sub-distributions of $\Prob$. The figure demonstrate, the test queries are evenly from all the  sub-distributions, where the training query workload is composed by skew distributed sub-distributions.
}

\stitle{Scope and limitations.}  We concentrate on the \emph{covariant shift} problem of DL-based query-driven cardinality estimation, where
the distribution of the test query workload shifts from the distribution of the training workload regarding the \emph{query features}.
In other words, the algorithms we explore aim to improve the overall generalization capability when the distribution of the test query features shifts from that of the training query features. 
In this paper, the algorithms we study are not specifically designed to deal with the \emph{label shift} problem, which is incurred by database updates and the shifting derives from changes in the true cardinality.
In addition, the algorithms we study are offline training algorithms that build one-off models from a stationary training query workload. Online learning approaches with a dynamic training query workload, e.g., active learning and incremental learning~\cite{DBLP:conf/sigmod/LiLK22, DBLP:conf/sigmod/ZhaoYHLZ22, DBLP:journals/pacmmod/WuI24}, are orthogonal to our research, which can be used to further improve the model based on specific system design choices.
Recent techniques involving enhancing the query representation by sampling-based features~\cite{DBLP:journals/pvldb/NegiWKTMMKA23} and model ensemble~\cite{DBLP:journals/pvldb/LiuD0Z21} are also parallel to our study. 

\comment{
\begin{table}[t]
{\small 
\caption{Frequented-used Notations}
\label{tbl:implem}
\vspace*{-0.4cm}
\begin{center}
    \begin{tabular}{c c} \toprule
     {Notation} & { Description } \\ \midrule
   $q$ & SQL query \\  \midrule           
   $c(q)$ / $c_\Theta(q)$ & true / estimated cardinality of query $q$ \\ \midrule
    & \\
   \bottomrule
   \end{tabular}
\end{center}
}
\end{table}
}

\section{Methodology}
\label{sec:method}

We categorize the learning algorithms we explore into three classes: representation learning, data manipulation, and new learning strategies. We propose a novel learning algorithm by self-supervised learning, which is in the category of learning strategy.

\comment{
\begin{algorithm}[t]
\small
\SetAlgoLined
\SetKwRepeat{Do}{do}{while}
\KwIn{Training query set $\mathcal{D}=\{(q_i, c(q_i))\}_{i = 1}^{|\mathcal{D}|}$, learning rate $\eta$;
}
\KwOut{Model ${\Theta}$}
Randomly initialize parameter $\Theta$;\\
\Do{$\mse$ converge}{
Sample a batch $\batch \sim \mathcal{D}$; \\
Compute the loss $\mse(\batch, \Theta)$ via Eq.~\eqref{eq:loss:erm}; \\
Update the model $\Theta \leftarrow \Theta - \eta \nabla_{\Theta} \mse(\batch, \Theta)$;
}
\caption{Empirical Risk Minimization}\label{alg:erm}
\end{algorithm}
}

\subsection{Representation Learning}
The representation learning strategy aims to learn an invariant query embedding under  distribution shift. We decompose the learned estimator $c_{\Theta}$ as $c_{\Theta} = g_{\Theta} \circ h_{\Theta}$, where $h_{\Theta}$ is the feature extraction function that transforms an initial query encoding into an intermediate vector embedding, and $g_{\Theta}$ is the prediction function that predicts the cardinality from the intermediate embedding. The goal of the representation learning strategy can be formulated as Eq.~\eqref{eq:represent}, where $\loss_{reg}$ denotes a regularization term and $\lambda$ is the trade-off parameter.
\begin{align}
\min_{h_{\Theta}, g_{\Theta}} \mathbb{E}_{q, c(q)}[ \norm{ g_{\Theta} \circ h_{\Theta}(q) - \log c(q)}^2 ]  + \lambda \loss_{reg} \label{eq:represent}
\end{align}
The representation learning strategy is designed to better learn the 
feature extraction function $h_{\Theta}$ with corresponding $\loss_{reg} $.

\begin{algorithm}[t]
\small
\SetAlgoLined
\SetKwRepeat{Do}{do}{while}
\KwIn{Training query set $\mathcal{D}=\{(q_i, c(q_i))\}_{i = 1}^{|\mathcal{D}|}$, learning rate $\eta$;
}
Randomly initialize parameter $\Theta$;\\
Partition queries in $\mathcal{D}$ to groups $\mathcal{D}_i, \cdots, \mathcal{D}_m$; \\
Compute the sampling weights $p_i \leftarrow {|\mathcal{D}_i|}/{\sum_{i=1}^{m} |\mathcal{D}_i|}$; \\
\Do{$\loss$ converge}{
Sample two groups $\mathcal{D}_i, \mathcal{D}_j$ by weights $p = (p_1, \cdots, p_m)$; \\
Sample two batches $\batch_i \sim \mathcal{D}_i, \batch_j \sim \mathcal{D}_j$;\\
Compute the loss $\loss \leftarrow \mse(\batch_i; \Theta) + \mse(\batch_j; \Theta) + \lambda \loss_{align}(\batch_i, \batch_j; \Theta)$; \\
Update the model $\Theta \leftarrow \Theta - \eta \nabla_{\Theta} \loss$;
}
\caption{Deep Correlation Alignment}\label{alg:coral}
\end{algorithm}

\stitle{Deep Correlation Alignment (Deep CORAL)} \cite{DBLP:conf/eccv/SunS16} is a classical transfer learning algorithm to alleviate distribution shift. It is originally designed to adapt the neural networks trained from a source domain to a target domain, given the unsupervised data from the target domain. We extend Deep CORAL to solve the OOD generalization problem under query distribution shift, where the target query distribution is unknown in the training stage.
Suppose the training query set $\Dataset$ can be divided into different groups $\{\Dataset_1, \cdots \Dataset_m\}$, e.g., by the number of selection and join predicates, or query templates,  where each group is subject to a sub-distribution, $\Dataset_i \sim \Prob_i$. Deep CORAL learns a query feature extractor $h_\Theta : q \mapsto
\mathbb{R}^d$ that aligns the feature correlation across these sub-distributions. 
Specifically, Eq.~\eqref{eq:loss:align} defines the regularization term $\loss_{reg}$ of Deep CORAL, which minimizes the distance of the covariance matrices of the feature between any two groups $\Dataset_i$ and $\Dataset_j$. Here, $\bm{C}_{\Dataset_i} \in \mathbb{R}^{d \times d}$ denotes the covariance matrix of the query feature in the group $\Dataset_i$, and the query feature embedding is computed by $h_\Theta$. 
\begin{align}
\loss_{align}(\Dataset_i, \Dataset_j; \Theta) &= \sum_{i = 1}^{m} \sum_{j = 1}^{m} \frac{1}{4d^2} \norm{\bm{C}_{\Dataset_i} - \bm{C}_{\Dataset_j}}^2 \label{eq:loss:align} 
\end{align}

Algorithm~\ref{alg:coral} presents the training algorithm of Deep CORAL, which jointly optimizes the MSE loss and Eq.~\eqref{eq:loss:align} by mini-batch SGD.
Given the $m$ partitioned training query groups, in each step, we sample two groups $\Dataset_i, \Dataset_j$ by the weights which are propositional to the group size. The weighted sampling enables queries in different sizes of groups can be used in the training uniformly. 
Then, we sample two mini-batches with an equal batch size from the two groups, respectively. We compute the MSE loss plus the regularization loss of Eq.~\eqref{eq:loss:align} for the two batches as the final loss.

\stitle{Unsupervised Domain Adaptation (DANN)}  \cite{DBLP:conf/icml/GaninL15} promotes the feature extractor $h_{\Theta}$ to generate invariant embedding w.r.t. distribution shift implicitly by adversarial training, instead of explicit feature alignment like Deep CORAL. Apart from the decomposition of $c_{\Theta} = g_{\Theta} \circ h_{\Theta}$ for the regression task, DANN introduces an additional discriminator  $y_\Theta(q) = [\hat{p}(\Dataset_i, q) | i = 1,\cdots, m]$ which predicates the group of a query by taking the query embedding $h_{\Theta}(q)$ as the input. Here, $\hat{p}(\Dataset_i, q) \in [0, 1]$ denotes the probability that query $q$ is in the group $\Dataset_i$.
The objective of DANN is to learn a feature extractor $h_{\Theta}$ 
to make the discriminator indistinguishable for the groups as far as possible, while keeping the effectiveness of the main regression task. Thereby, the negation of the classification loss, i.e., the cross entropy (CE) loss of Eq.~\eqref{eq:loss:cla} serves as the regularization term $\loss_{reg}$, where $p(\Dataset_i, q) \in \{ 0, 1\}$ is the binary label indicating whether $q$ is in the group $\Dataset_i$.
\begin{align}
\mathcal{L}_{ce}(\Dataset; \Theta) &= \sum_{q \in \mathcal{D}} \sum_{i = 1}^m [- p(\Dataset_i, q) \log \hat{p}(\Dataset_i, q)] \label{eq:loss:cla}
\end{align}

Algorithm~\ref{alg:dann} shows the training procedure of DANN we extend. We first assign a class label $g_i$ for each training query $q_i$ given its derived sub-distribution. In each gradient step, the algorithm minimizes the MSE loss $\loss_{mse}$ and maximizes the CE loss $\loss_{ce}$ simultaneously, where the corresponding parameters of the feature extractor $\Theta_{h}$, the regression layer $\Theta_{c}$ and the classification layer $\Theta_{g}$ are updated.

\begin{algorithm}[t]
\small
\SetAlgoLined
\SetKwRepeat{Do}{do}{while}
\KwIn{Training query set $\mathcal{D}=\{(q_i, c(q_i))\}_{i = 1}^{|\mathcal{D}|}$, learning rate $\eta$;
}
Randomly initialize parameter $\Theta$;\\
Add group label to training set $\mathcal{D} \leftarrow \{(q_i, c(q_i), g_i)\}_{i = 1}^{|\mathcal{D}|}$; \\
\Do{$\mse$ converge}{
Sample a batch $\batch \sim \mathcal{D}$; \\
Compute the loss $\mse(\batch;  {\Theta_h}, {\Theta_c})$ via Eq.~\eqref{eq:loss:erm} and $\loss_{ce}(\batch, {\Theta_h}, {\Theta_g})$ via Eq.~\eqref{eq:loss:cla}; \\
$\Theta_h \leftarrow \Theta_h - \eta (\nabla_{\Theta_h} \mse(\batch; {\Theta_h}, {\Theta_c}) +\nabla_{\Theta_h}\loss_{ce}(\batch; {\Theta_h}, {\Theta_g}))$;\\
$\Theta_c \leftarrow \Theta_c - \eta \nabla_{\Theta_c} \mse(\batch; {\Theta_h}, {\Theta_c})$ ;\\
$\Theta_g \leftarrow \Theta_g + \eta \nabla_{\Theta_g} \loss_{ce}(\batch; {\Theta_h}, {\Theta_g})$ ;\\
}
\caption{Unsupervised Domain Adaptation}\label{alg:dann}
\end{algorithm}

\subsection{Learning Strategy}
General ML strategies can also be used to alleviate model degradation under distribution shift, which can be divided into the categories of ensemble learning, meta-learning, Group Distributionally Robust Optimization (Group DRO) and self-supervised learning, etc. We explore the Group DRO algorithm that achieves superb empirical performance. Afterwards, we design a new self-supervised learning algorithm that leverages the partial order of SQL queries in a self-supervised learning task. 

\stitle{Group Distributionally Robust Optimization (Group DRO)} \cite{DBLP:journals/corr/abs-1911-08731} is to learn a model at a worst-case scenario to envision it can generalize to test data.
Given the training queries that are divided into $m$ groups $\{\Dataset_1, \cdots, \Dataset_m\}$, where $\Dataset_i \sim \Prob_i$, it assumes the test distributions is an uncertainty set that can be any mixture of the $m$ sub-distributions, i.e., $\mathcal{Q} := \{\sum_{i = 1}^m \omega_i \Prob_i | \bm{\omega} \in \Delta_{m} \}$, where $\Delta_{m}$ is $(m-1)$-dim probability simplex.
Group DRO minimizes the worst-group empirical loss in Eq.~\eqref{eq:loss:worst-case-exp}, and hopes to generalize over the uncertainty set of distributions $\mathcal{Q}$.
\begin{align}
    \loss = \max_{\Prob_i} \loss_{exp}(\Prob_i; \Theta) \label{eq:loss:worst-case-exp}
\end{align}
To train Group DRO models efficiently, an SGD-based  algorithm is proposed with convergence guarantees. The algorithm minimizes an equivalent form of Eq.~\eqref{eq:loss:worst-case-exp}, i.e., Eq.~\eqref{eq:loss:dro}, which is shown in Algorithm~\ref{alg:dro}.
\begin{align}
\mathcal{L}_{dro} = \sup_{\bm{\omega} \in \Delta_{m}} \sum_{i = 1}^{m} \omega_i \loss_{exp}(\Prob_i; \Theta) \label{eq:loss:dro}
\end{align}
Intuitively, Algorithm~\ref{alg:dro} maintains a distribution $\bm{\omega}$ over groups, with high masses on high-loss groups, and updates on each training query proportionally to the mass on its group. First, $\bm{\omega}$ is sampled from a Uniform distribution. Then, the algorithm interleaves SGD on the model parameter $\Theta$ and exponential gradient ascent on $\bm{\omega}$, where the two parameters are updated mutually.  

\begin{algorithm}[t]
\small
\SetKwRepeat{Do}{do}{while}
\KwIn{Training query set $\mathcal{D}=\{(q_i, c(q_i))\}_{i = 1}^{|\mathcal{D}|}$, \\
learning rates $\eta_{\Theta}$, $\eta_{\omega}$;
}
Randomly initialize parameter $\Theta$; \\
Initialize $\bm{\omega} \sim \text{Uniform}(1, \cdots m)$; \\
\Do{$\mse$ converge}{
Sample a batch $\batch \sim \mathcal{D}$; \\
Create group-level empty batches $\batch_1, \cdots, \batch_m \leftarrow \emptyset, \cdots, \emptyset$; \\
\For{$(q_i, c(q_i)) \in \batch$}{
    Assign $(q_i, c(q_i))$ to corresponding batch $\batch_j$; \\
    }
\For{$i \in [1, \cdots, m ]$}{
    Compute group-level loss $\mse(\batch_i; \Theta)$ via Eq.~\eqref{eq:loss:erm};\\
    Compute new weight $\omega'_i \leftarrow \omega_i \exp(\eta_{\omega} \mse(\batch_i; \Theta))$;\\
}
Update the weights $\omega_i \leftarrow {\omega'_i}/{\sum_{i=1}^{m} \omega'_i}$; \\
Update the model $\Theta \leftarrow \Theta - \eta_{\Theta} \nabla_{\Theta} \sum_{i = 1}^{m}\mse(\batch_i; \Theta)$;
}
\caption{Group Distributionally Robust Optimization}\label{alg:dro}
\end{algorithm}

\stitle{Self-Supervision with Order Embedding (\OrderEmb)}.
Self-supervised learning (SSL)~\cite{DBLP:journals/pami/JingT21} is a recently popular learning paradigm that exploits self-supervised auxiliary tasks from large-scale unlabeled data.
In computer vision, SSL is used to learn generalized representations to alleviate the distribution shift from training to test domain. Various computer vision tasks that are relevant to the main learning tasks (e.g., image classification) are proposed as the SSL tasks, including solving jigsaw puzzles~\cite{DBLP:conf/cvpr/CarlucciDBCT19}, rotation angle prediction~\cite{DBLP:conf/icml/SunWLMEH20}, and in-class feature similarity comparison~\cite{DBLP:conf/iccv/KimYPKL21}.

We specifically design an SSL task to support query distribution shift in the context of cardinality estimation. 
The gist of our approach is a contrastive learning strategy that restricts the partial order relations of query conditions are properly reflected in the query embedding space: if the conjunctive condition of a query $q'$ is a sub-condition of a query $q$, indicating there is a partial order relation in the true cardinality of the two queries, $c(q') \le c(q)$, we wish the embedding $h_{\Theta}(q')$ will be on the `lower-left' position of the embedding $h_{\Theta}(q)$ in the query embedding space, i.e., $h_{\Theta}(q')[i] \le h_{\Theta}(q)[i], \forall i \in [1 \cdots, d]$. The embedding constraint is formulated by a contrastive loss on the query embeddings, as shown in Eq.~\eqref{eq:loss:order}, where $\mathcal{C}(q)$ denotes a set of contrastive queries of $q$, in which each query is subject to the partial order sub-condition with $q$.  
\begin{align}
\mathcal{L}_{ord} &=  \sum_{q \in \mathcal{D}} \sum_{q' \in \mathcal{C}(q)} \max \{ \norm{ h_{\Theta}(q') - h_{\Theta}(q)},0\} \label{eq:loss:order} 
\end{align}
The contrastive loss, serving as a regularization term, is jointly optimized with the MSE loss of the main task, and the training procedure is shown in Algorithm~\ref{alg:order}. 
In each gradient step, when a batch of queries is sampled, we first sample $k$ contrastive queries for 
each training query $q$ in the current batch. 
Concretely, by taking query $q$ as a template, we uniformly draw some selection conditions and revise these sampled selection conditions to generate a query $q'$. 
For numerical range conditions, we uniformly sample an intermediate value from the query range, then shorten the query range by setting one of the start/end points as the intermediate value. 
For categorical \IN filter conditions, we uniformly remove some elements from the filter set.
This sampling strategy ensures that the true cardinality of $q'$ is no larger than that of $q$, i.e., $c(q) \le c(q')$.
It is worth noting that our SSL-based approach does not require each training query to possess an explicit group label, thereby this approach is more flexible when the underlying sub-distributions $\{\Prob_1, \cdots, \Prob_m\}$ are hard to depict. 
\begin{algorithm}[t]
\small
\SetAlgoLined
\SetKwRepeat{Do}{do}{while}
\KwIn{Training query set $\mathcal{D}=\{(q_i, c(q_i))\}_{i = 1}^{|\mathcal{D}|}$, learning rate $\eta$;
}
Randomly initialize parameter $\Theta$;\\
\Do{$\loss$ converge}{
Sample a batch $\batch \sim \mathcal{D}$; \\
\For{$(q_i, c(q_i)) \in \batch$}{
    Sample $k$ contrastive queries for $q_i$ as set $\mathcal{C}(q_i)$;\\
    }
Compute the loss $\loss \leftarrow \mse(\batch; \Theta) + \lambda \loss_{ord}(\batch; \Theta)$; \\
Update the model $\Theta \leftarrow \Theta - \eta \nabla_{\Theta} \loss(\batch; \Theta)$;
}
\caption{Self-supervision with Order Embedding}\label{alg:order}
\end{algorithm}

\subsection{Data Manipulation}
ML models are always hungry for more training data. Given a limited set of training data, data manipulation is the simplest and cheapest way to enhance the diversity of the training data, in that to enhance the generalization capability of the models. Let $\tilde{q},c(\tilde{q}) = \mathcal{M}(\tilde{q},c(\tilde{q}))$ denote the manipulated data using a function $\mathcal{M}(\cdot)$, we formulate the general learning objective of data manipulation-based approaches for OOD generalization as the Eq.~\eqref{eq:loss:data} below:
\begin{align}
 \mathbb{E}_{q, c(q)} \norm{\log c(q) - \log c_\Theta(q)}^2 + \mathbb{E}_{\tilde{q}, c(\tilde{q})} \norm{\log c(\tilde{q}) - \log c_\Theta(\tilde{q})}^2, \label{eq:loss:data}
\end{align}
where different approaches adopt different functions $\mathcal{M}(\cdot)$.

\stitle{Query Mixup}.
We extend a popular data generation technique, Mixup~\cite{DBLP:conf/iclr/ZhangCDL18} for learning query-driven cardinality estimators, which has achieved promising results for improving the generalization under distribution shift in image, text, and graph domains~\cite{DBLP:conf/icml/HanJLH22, DBLP:conf/coling/SunXYLYH20}. As shown in Eq.~\eqref{eq:mixup:mix}, Mixup generates new training data by performing linear interpolation between two instances and between their labels, between a weight sampled from a Beta distribution, i.e., $\xi \sim \text{Beta}(\alpha, \alpha)$. Here, $e(q)$ denotes the initial encoding of a query $q$. The interpolated data $(e(\tilde{q}), c(\tilde{q}))$ will be used as the augmented data to train the model.
\begin{align}
e(\tilde{q}) &= \xi \cdot e(q_i) + (1 - \xi) \cdot e(q_j) \nonumber \\
c(\tilde{q}) &= \xi \cdot c(q_i) + (1 - \xi) \cdot c(q_j) \label{eq:mixup:mix}
\end{align}
Instead of uniformly sampling two instances for interpolation, for regression tasks like cardinality estimation, we follow the C-Mixup~\cite{DBLP:conf/nips/YaoWZZF22} to sample two instances based on their label similarity.  
Given one query $q_i$, a Gaussian kernel is employed to compute the sampling probability of drawing another query $q_j$ for mixing, where queries with closer true cardinalities to $q_i$ are more likely to be sampled.
\begin{align}
    \bm{P}[q_i, q_j] \propto \exp (-\frac{\norm{\log c(q_i) - \log c(q_j)}^2}{\sigma^2}) \label{eq:mixup:dist}
\end{align}
In reality, linear interpolation will generate queries that violate the accurate query semantics, e.g., linear interpolation on  one-hot encoding of the selections, joins, and relations.

\begin{algorithm}[t]
\small
\SetAlgoLined
\SetKwRepeat{Do}{do}{while}
\KwIn{Training query set $\mathcal{D}=\{(q_i, c(q_i))\}_{i = 1}^{|\mathcal{D}|}$, learning rate $\eta$, hyper-parameter $\alpha$, Gaussian bandwidth $\sigma$;
}
Randomly initialize parameter $\Theta$;\\
Compute pair-wise distance matrix $\bm{P}$ for $q_i, q_j \in \mathcal{D}$ via Eq.~\eqref{eq:mixup:dist}; \\
\Do{$\mse$ converge}{
Sample a batch $\batch \sim \mathcal{D}$; \\
Create an empty batch as $\tilde{\batch} \leftarrow \emptyset$; \\
    \For{$(q_i, c(q_i)) \in \mathcal{B}$}{
    Sample a query $(q_j, c(q_j)) \sim \mathcal{D}$ by weight $\bm{P}[q_i, :]$;\\
    Sample $\xi \sim \text{Beta}(\alpha, \alpha)$;\\
    Compute $(\tilde{q}, c(\tilde{q}))$ via Eq.~\eqref{eq:mixup:mix} and insert to  $\tilde{\batch}$ ;\\
    }
Compute loss $\mse(\tilde{\batch}; \Theta)$ via Eq.~\eqref{eq:loss:erm} on $\tilde{\batch}$; \\
Update the model $\Theta \leftarrow \Theta - \eta \nabla_{\Theta} \mse(\tilde{\batch}; \Theta)$;
}

\caption{Query Mixup}\label{alg:mixup}
\end{algorithm}

\stitle{Query Masking} \cite{DBLP:journals/pvldb/NegiWKTMMKA23} is a recently proposed training strategy for robust query-driven cardinality estimation.
During the training stage, Query Masking uniformly drops selection predicates with a probability of $p$, where the corresponding features are masked as zero. In inference stage, full query features are fed into the model. Query masking aims to force the model to learn a `harder' task, i.e., predicting the cardinalities with missing query features.  

\section{{\CardOOD} Framework}
\label{sec:cardood}

In this section, we elaborate on the \CardOOD framework we establish, which integrates the 6 ML algorithms presented in \cref{sec:method} as well as ERM training strategy.
The framework contains 4 abstract modules, \emph{Query Encoder}, \emph{Training Pipeline}, \emph{Model} and \emph{Algorithm}, which are incorporated at a bottom-up level.
All the 4 modules are extensible by implementing plug-and-play instances. 
\emph{Query Encoder} is the module to transform raw SQL query statements into vectorized representations, which depends on the model type. Here, we implement a fixed-length encoder and set encoder for constructing the input vector representation for two representative query-driven models, Multi-layer Perceptron (MLP) and Multi-set Convolutional Network (MSCN), respectively.

\stitle{Fixed-length Encoding}:
In existing work~\cite{DBLP:journals/pvldb/DuttWNKNC19, dutt2020efficiently}, a
SPJ query can be encoded by a fixed-length vector. The encoding consists of two parts: the selection conditions
and the join conditions. The two parts are encoded separately and
concatenated. 
The selection conditions are specified on numerical/categorical
attributes.  For a range filter, $lb_j \leq A_j \leq ub_j$, on a
numerical attribute $A_j$, we normalize $lb_j$ and $ub_j$ to $[0, 1]$
by mapping $[lb_j, ub_j]$ to $[\frac{lb_j - min_j}{max_j - min_j},
  \frac{ub_j - min_j}{max_j - min_j}]$, where $[min_j, max_j]$ is the
domain of the attribute $A_j$. Thus, the representation is the
interval of two real values.
The equality predicates of categorical attributes are treated as range filters where $lb_j = ub_j$~\cite{DBLP:journals/corr/abs-2012-06743}.
For an \IN~filter, $A_j~\IN~C$, on a categorical attribute $A_j$,
where $C$ is a subset of the attribute $A_j$'s domain,
we adopt the lossless factorized bitmap encoding~\cite{DBLP:journals/corr/abs-2006-08109}. 
Regarding join conditions, we use an $n$-dim binary vector to represent equi-joins of total $n$ joinable attribute pairs.

\stitle{Set Encoding}: Following~\cite{DBLP:conf/cidr/KipfKRLBK19}, we encode a SPJ query into 3 sets of vectors, i.e., a set of vectors of  involved relations, selection conditions and join conditions, respectively. 
The 3 sets of vector are processed by set convolution, followed by average pooling layers. Subsequently, the output individual set representations are concatenated and fed into a final MLP. 

\begin{table}[t]
{\small 
\caption{Characteristics of \CardOOD}
\label{tbl:summary}
\begin{center}
    \begin{tabular}{c c c c c} \toprule
     {Algorithm} & \makecell[c]{Group \\ labels}   & \makecell[c]{OOD \\ Generalization} & \makecell[c]{Training \\ efficiency} & \makecell[c]{ Trainability}
    \\ \midrule                                            
   ERM & \ding{55}   & \star  & \star\star\star & \star\star\star   \\ 
   Deep CORAL & \ding{51}  & \star\star & \star &  \star\star   \\
   DANN & \ding{51} & \star\star & \star\star & \star  \\
   Group DRO & \ding{51}  & \star & \star\star & \star  \\
   OrderEmb & \ding{55}  & \star\star\star & \star & \star\star\star  \\
   Query Mixup & \ding{55}  & \star\star\star & \star\star & \star\star\star  \\
   Query Masking & \ding{55}  & \star & \star\star\star & \star\star\star  \\
   \bottomrule
   \end{tabular}
\end{center}
}
\end{table}

\comment{
\begin{table}[t]
{\small 
\caption{\CardOOD Implementation}
\label{tbl:implem}
\vspace*{-0.4cm}
\begin{center}
    \begin{tabular}{c c} \toprule
     {Modules} & { Implemented Instances }
    \\ \midrule                                            
   \cellcolor{LightRed}  \emph{Algorithm} &   {\makecell[c]{ERM, \\ Deep CORAL, DANN, \\ Group DRO, OrderEmb, \\ Query Mixup, Query Masking}} \\ \midrule
   \cellcolor{LightCyan} \emph{Model} &  \makecell[c]{Multi-layer Perceptron (MLP),\\ Multi-set Convolutional Network (MSCN)} \\ \midrule
   \cellcolor{LightYellow} \emph{Training Pipeline} &  Group-based, Sampling-based, Mask-based \\ \midrule
   \cellcolor{LightGray} \emph{Query Encoder} & Fixed-length Encoder, Set Encoder \\
   \bottomrule
   \end{tabular}
\end{center}
}
\end{table}
} 

\comment{
\stitle{Query Encoding}: Following the existing work, a
select-project-join query we support can be encoded by a fixed length
vector. The encoding consists of two parts: the selection conditions
and the join conditions. The two parts are encoded separately and
concatenated as follows. 
\[
\Scale[1.0]{
\overbrace{\underbrace{~0.25~0.5~}_{R_1.A_1}~\cdots~}^{R_1~\text{selections}}\overbrace{\underbrace{~1~1~0~1~0~}_{R_2.A_2}~\cdots~}^{R_2~ \text{selections}}\overbrace{\underbrace{~1~1~0~}_{(R_1.A_3, R_2.A_3)}~0~0~0~
}^{\text{ joins}}
}
\]
For selection condition, the encoding of all the attributes in all the
relations of the schema are concatenated by a fixed order (e.g.,
lexicographical order).  In a similar manner, for the join conditions,
the encoding of all the join pairs are concatenated.

The selection conditions are specified on numerical/categorical
attributes.  For a range filter, $lb_j \leq A_j \leq ub_j$, on a
numerical attribute $A_j$, we normalize $lb_j$ and $ub_j$ to $[0, 1]$
by mapping $[lb_j, ub_j]$ to $[\frac{lb_j - min_j}{max_j - min_j},
  \frac{ub_j - min_j}{max_j - min_j}]$, where $[min_j, max_j]$ is the
domain of the attribute $A_j$. Thus, the representation is the
interval of two real values.
The equality predicates of categorical attributes are treated as range filter where $lb_j = ub_j$~\cite{DBLP:journals/corr/abs-2012-06743}.
For an \IN~filter, $A_j~\IN~C$, on a categorical attribute $A_j$,
where $C$ is a subset of the attribute $A_j$'s domain $\{c_1, c_2,
\dots c_m\}$, a straightforward encoding is to build an $m$-dim bitmap
where its $k$-th bit is 1 if $c_k \in C$, otherwise 0.  This binary
representation is effective for attributes with small domain, however,
it is difficult to scale on a large domain where the predicate vector
is high-dimensional and sparse.  Therefore, for a large domain, we
adopt the factorized bitmap~\cite{DBLP:journals/corr/abs-2006-08109},
i.e., slicing the whole bitmap to chunks, and converting each chunk
into corresponding base-10 integer. Finally, the selection condition
is represented losslessly by $\lceil m / s \rceil$ integers, where $s$
is the length of the chunk.
}

\comment{
Regarding join conditions, for each joinable attribute pair $(A_i,
A_j)$, we use a 3-bit bit-map to encode the join condition on this
pair, corresponding to the 3 comparison operators, $<, =, >$,
respectively, where `1' denotes there is a comparative condition on
$(A_i, A_j)$. For example, $A_i < A_j$, $A_i \geq A_j$, and $A_i \neq
A_j$ are encoded as `100', `011' and `101', respectively. The bit-map
`000' denotes that the query is free of join condition on the pair.
}

\emph{Training Pipeline} is a module that further reorganizes the vectorized training queries into mini-batches, forming the input stream for the learning algorithms. Thereby, \emph{Training Pipeline} is orthogonal to any model choice but depends on instances of \emph{Algorithm}. 
In \CardOOD, we build a Group-based pipeline that automatically partitions training queries into groups as training mini-batches or assigns explicit group labels for batched queries, which supports the algorithms Deep CORAL, DANN, and Group DRO. 
The sampling-based pipeline samples additional queries to construct a mini-batch, which are the contrastive queries for \OrderEmb or the interpolated queries for Query Mixup.
The masking-based pipeline randomly masks query features to support the algorithm Query Masking.

We summarize the characteristics of the 6 ML algorithms as well as the baseline ERM, from the perspective of the requirement of group labels, the OOD generalization capability, training efficiency, and the trainability (encompassing the stability of training, convergence rate, and sensitivity to hyper-parameters). Table~\ref{tbl:summary} lists a general assessment of the 7  algorithms for the characteristics. For group labels, 3 algorithms, Deep CORAL, DANN, and Group DRO, need to partition the training queries into groups w.r.t. some sub-distributions, where DANN needs explicit group labels for loss computation. 
For OOD generalization, Deep CORAL and OrderEmb
achieve the overall best performance in single relation queries,
while Query Mixup and OrderEmb achieve the best performance
in multi-join queries. 
For training efficiency, taking ERM as the baseline, Query Masking is fast to train a model, while DANN, Group DRO, and Query Mixup are less efficient. 
Deep CORAL and OrderEmb are the worst-efficient algorithms for training, due to the large overhead of computing the covariance matrix and sampling contrastive queries online, respectively. 
For trainability, in our testing, DANN and Group are relatively difficult to train. Concretely, DANN introduces an adversarial optimization objective and Group DRO needs to optimize model parameters and group parameters mutually.
Deep CORAL needs a little effort for hyperparameter tuning. 
For the other 4 algorithms, their training dynamics and model performance are robust to hyper-parameters in the empirical ranges.

\section{Experimental Studies}
\label{sec:exp}

\input{exp}

\section{Related Work}
\label{sec:rw}
\stitle{Classical Cardinality Estimators.}
Cardinality and selectivity estimation of relational queries have been studied in the database area for several decades.
In the early approaches, multi-dimensional histograms are proposed to represent the joint probability distribution~\cite{DBLP:conf/vldb/PoosalaI97, DBLP:journals/vldb/GunopulosKTD05, DBLP:conf/sigmod/PoosalaIHS96}, where the assumption of attribute value independence is not necessary to be held. 
For join queries, various join sampling algorithms~\cite{haas1999ripple, DBLP:conf/sigmod/0001WYZ16, DBLP:conf/sigmod/CaiBS19, DBLP:conf/sigmod/ZhaoC0HY18} are developed, which can be used to approximate cardinality and query results. 
However, these approaches face the risk of sampling failure in the intermediate join when the data distribution is complex. 
Recent work by
\cite{DBLP:conf/sigmod/GetoorTK01, DBLP:journals/pvldb/TzoumasDJ11} use Bayesian Network to estimate cardinality by modeling the joint probability distribution of all attributes.

\stitle{ML/DL for Cardinality Estimation and AQP.}
ML/DL models are exploited to perform cardinality estimation and approximate query processing (AQP)
for \rdbm.  
These approaches are categorized into supervised learning approaches and unsupervised learning approaches. Supervised learning approaches are query-driven, which learn a function that maps the features of a query to its cardinality. Various ML/DL models, such as Gradient Boosting Decision Tree (GBDT)~\cite{DBLP:journals/pvldb/DuttWNKNC19, dutt2020efficiently}, Fully-connected Neural Networks~\cite{DBLP:journals/pvldb/LiuD0Z21, DBLP:journals/pvldb/DuttWNKNC19}, Multi-set Convolutional Network (MSCN)~\cite{DBLP:conf/cidr/KipfKRLBK19, DBLP:journals/pvldb/NegiWKTMMKA23}, serve as the functions to be learned. Unsupervised learning approaches are data-driven, which learn the joint probability distribution of the underlying relational data. 
Deep models including Sum-Product Network~\cite{DBLP:journals/pvldb/HilprechtSKMKB20}, Autoencoders~\cite{DBLP:conf/icde/Thirumuruganathan20, DBLP:journals/pvldb/YangLKWDCAHKS19,
  DBLP:conf/sigmod/HasanTAK020}, and Transformer~\cite{DBLP:journals/pvldb/YangLKWDCAHKS19} are used to model the joint probability distribution, where cardinality queries are compiled to probability queries or sampling queries on the models. 
Supervised learning approaches are easy to deploy with relatively low training and prediction overhead, however, they lack robustness to shifting query workloads. On the contrary, unsupervised learning approaches are robust to different query workloads. However, model construction consumes many resources as large volumes of data need to be fed into the model multiple times. 
\comment{
\DBEst~\cite{DBLP:conf/sigmod/MaT19} builds kernel density
estimation (KDE) models and tree-based regressors to conduct AQP.
\DeepDB~\cite{DBLP:journals/pvldb/HilprechtSKMKB20} adopts Sum-Product
Networks (SPNs) to learn the joint probability distribution of
attributes. An SQL query is compiled to a product of expectations or
probability queries on the SPNs, where the product is based on the 
independence of attributes.
\cite{DBLP:conf/icde/Thirumuruganathan20} uses deep generative models,
e.g., Variational Autoencoder (VAE) to model the joint probability
distribution of attributes, which supports analytical aggregate
queries on a single relation.  \cite{DBLP:journals/pvldb/KieferHBM17}
estimates multivariate probability distributions of a relation to
perform cardinality estimation by kernel density estimation (KDE).  
Deep autoregressive models, e.g.,
Masked Autoencoder (MADE), are also adopted to learn the joint
probability distribution \cite{DBLP:journals/pvldb/YangLKWDCAHKS19,
  DBLP:conf/sigmod/HasanTAK020}, which decompose the joint
distribution to a product of conditional distributions.  Kipf
et al. propose a multi-set convolutional neural network (MSCN) to
express query features using sets \cite{DBLP:conf/cidr/KipfKRLBK19}.
Anshuman et al.~\cite{DBLP:journals/pvldb/DuttWNKNC19} use
MLP and tree-based regressor to express
multivariate range queries.  Sun
et al.~\cite{DBLP:journals/pvldb/SunL19} extract the features of
physical query plan by Tree LSTM to estimate the query execution cost
as well as the cardinality.
\Fauce~\cite{DBLP:journals/pvldb/LiuD0Z21} adopts \DeepEnsemble~\cite{DBLP:conf/nips/Lakshminarayanan17}, an ensemble of fully-connected neural networks, where each network predicts the mean and variance of a Gaussian distribution.
Two types of uncertainties, model uncertainty and data uncertainty are derived from the predictive distributions~\cite{DBLP:conf/aaai/XiaoW19}.
}

\stitle{Generalization Under Distribution Shift.} 
Modern ML has witnessed the challenges of dealing with non-\emph{i.i.d.} training and test datasets in real applications. Literature surveys on generalization under distribution shift can be found in~\cite{DBLP:journals/corr/abs-2108-13624, DBLP:journals/tkde/WangLLOQLCZY23, DBLP:conf/emnlp/YangSRLWZL0F023}. 
Regarding ML/DL for cardinality estimation, distribution shift on query workloads and the underlying distribution of the relational data influence the prediction accuracy of the learned models. To improve the robustness of the model, Warper~\cite{DBLP:conf/sigmod/LiLK22} adopts a Generative Adversarial Network to generate additional queries to mimic a shifted workload, and picks useful queries to annotate their true cardinality for online model update. 
Online learning can alleviate shifts from query workloads and data updates.  
\cite{DBLP:journals/pvldb/NegiWKTMMKA23} is a query-driven robust cardinality estimation approach. It randomly masks table or column features in the query during model training, forcing the model to rely on the additional features of DBMS statistics which could be more robust. 
DDUp~\cite{DBLP:journals/pacmmod/KurmanjiT23} is oriented to OOD detection and generalization for data-driven models. It detects distribution shift over relational data by statistical hypothesis testing on the loss value of the model. For in-distribution data updates, the model is updated by fine-tuning. For OOD data update, the model is updated by teacher-student knowledge distillation.   
In the recent, \cite{DBLP:journals/pacmmod/KurmanjiTT24} investigates the maintenance of DL models in database applications, when databases face deletion operations. Several machine learning algorithms are studied for models of AQP, cardinality estimation, data generation, and data classification.  
Our framework is relevant but orthogonal to these studies, where we concentrate on establishing a robust query-driven learned estimator via one-off training. 


\section{Conclusion}
\label{sec:conclusion}
DL-based query-driven cardinality estimators face the challenge of distribution shift, where queries to be predicted are OOD from the training query workload. 
In this paper, we build a framework \CardOOD for training robust query-driven cardinality estimators. \CardOOD integrates 6 learning algorithms over two DL models. 
Among these algorithms, 4 algorithms are the extensions of the transfer/robust learning techniques in ML areas.  
In addition, we propose a new algorithm that leverages the partial order of the cardinality among training queries to improve model generalization in OOD scenarios.  
Our extensive experiments on \CardOOD show that the robust learned estimators achieve much lower  $95\%$ and $99\%$ quantiles of the \Qerror than the estimators trained by plain ERM training paradigm. 
We also deploy \CardOOD in a real DBMS, \PostgreSQL, which improves the speed of the end-to-end query execution by up to $5.6\%$, $36.6\%$ 
and $4.6\%$ for \imdb, \dsb and \job queries, respectively.

\section*{Acknowledgments}
This work was supported by the Research Grants Council of Hong Kong, China, No. 14205520.
Kangfei Zhao is supported by National Key Research and Development Program, No. 2023YFF0725101.


\bibliographystyle{ACM-Reference-Format}
\bibliography{sample}
\end{document}

%% file: command.tex
\newcommand{\stitle}[1]{\vspace{1ex} \noindent{\bf #1}}

\newcommand{\kw}[1]{{\ensuremath {\mathsf{#1}}}\xspace}

\newcommand*{\Scale}[2][4]{\scalebox{#1}{$#2$}}

\newcommand{\beqn}{\begin{eqnarray*}}
\newcommand{\eeqn}{\end{eqnarray*}}

\newcounter{ccc}

\newcommand{\kwnospace}[1]{{\ensuremath {\mathsf{#1}}}}

\newcommand{\PostgreSQL}{{\sl PostgreSQL}\xspace}

\newcommand{\DBEst}{{\sl DBEst}\xspace}
\newcommand{\DeepDB}{{\sl DeepDB}\xspace}

\newcommand{\Qerror}{\kwnospace{q}\mbox{-}\kw{error}}
\newcommand{\DeepEnsemble}{{\sl Deep~Ensemble}\xspace}

\newcommand{\Fauce}{{\sl Fauce}\xspace}

\newcommand{\OrderEmb}{{\text{OrderEmb}}\xspace}

\newcommand{\CardOOD}{{\sl CardOOD}\xspace}

\newcommand{\rdbm} {{\sl RDBMS}\xspace}
\newcommand{\PyTorch} {{\sl PyTorch}\xspace}
\newcommand{\Pandas} {{\sl Pandas}\xspace}

\newcommand{\IN}{\kw{IN}}
\newcommand{\OR}{\kw{OR}}

\newcommand{\forest}{\kw{forest}}

\newcommand{\imdb}{\kw{IMDB}}
\newcommand{\dsb}{\kw{DSB}}

\newcommand{\job}{\kwnospace{JOB}\mbox{-}\kw{light}}

\newcommand{\loss}{\mathcal{L}}
\newcommand{\mse}{\mathcal{L}_{mse}}
\newcommand{\batch}{\mathcal{B}}

\newcommand{\norm}[1]{\left\lVert#1\right\rVert}

\newcommand{\Model}{\mathcal{M}}
\newcommand{\Dataset}{\mathcal{D}}
\newcommand{\Prob}{\mathcal{P}}

\renewcommand{\star}{\ding{72}}

\newcommand{\ruimodify}[1]{\textcolor{blue}{#1}}

%% file: exp.tex
In this section, we give the test settings in~\cref{sec:exp:setup}, and report our comprehensive experiments in the following facets:
\ding{172} Test the prediction robustness of \CardOOD (\cref{sec:exp:robust})
\ding{173} Compare the training time of the algorithms in \CardOOD (\cref{sec:exp:efficiency})
\ding{174} Study the performance of order embedding regarding parameter sensitivity (\cref{sec:exp:sensitivity})
\ding{175} Investigate the utility of robust learned estimators for query optimization in \PostgreSQL (\cref{sec:exp:postgres}).

\subsection{Experimental Setup}
\label{sec:exp:setup}

\stitle{Datasets.} 
We use 3 relational datasets, one for single relation range queries, and two for multi-relation join queries. 
\forest~\cite{Dua:2019} originally contains 54 attributes of forest cover type. Following \cite{DBLP:journals/pvldb/DuttWNKNC19, DBLP:journals/vldb/GunopulosKTD05}, we use the first 10 numerical attributes. The relation has about 581K of rows. 
For \imdb \cite{DBLP:conf/job/Leis18}, we use 6 relations, including \textsf{title}, \textsf{cast\_info}, \textsf{movie\_info}, \textsf{movie\_companies}, \textsf{movie\_info\_idx}, \textsf{movie\_keyword}.
There are 18 attributes and 6 PK/FK join conditions in total. 
\dsb \cite{DBLP:conf/dsb/ding2021} builds upon the widely-used TPC-DS benchmark \cite{DBLP:conf/tpcds/Poess02} by adding intricate data distributions and challenging query templates. It contains 24 relations and includes both PK/FK joins and non-PK/FK many-to-many joins.

\stitle{Queries.}
We construct large query workloads in the following way.
For single relation \forest, we generate query sets with the number of select conditions varying from 2 to $D$ where $D$ is the number of attributes, and generate 2,000 queries for each subset. 
To generate a query of $d$ selection conditions, first, we uniformly sample $d$ attributes from all the $D$ attributes, then uniformly sample each attribute by the data-centric distribution following ~\cite{DBLP:journals/pvldb/DuttWNKNC19}. 
For multi-relation dataset \imdb, we test the query type supported by the existing baselines, i.e., PK/FK join without selection conditions on the join attributes. We generate query sets with the number of PK/FK joins, $t$, varying from 0 to $|T| - 1$, where $|T|$ is the number of involved relations.   
To generate a query of $t~(t > 0)$ joins, firstly a starting relation is uniformly sampled, then the query is constructed by traversing from the starting relation over the join graph in $t$ steps.
Here, for each relation of the sampled join query, additional selection conditions are drawn independently.  
For each $t$, 3,000 queries are generated for \imdb. 
Additionally, we generate a large query set of \job. 
\job~\cite{DBLP:conf/cidr/KipfKRLBK19} contains $70$ hand-written join 
queries that are suitable to serve as the query templates, as each query is different from others in involved set of relations and/or
selection conditions. We generate $1,200$ queries for each template by uniformly drawing the literals of the predicates in their corresponding range. 
\dsb \cite{DBLP:conf/dsb/ding2021} provides $15$ SPJ query templates for evaluating optimization techniques. We remove two templates, template 013 and 085, as they contain complex logical predicate \OR. 
For the rest $13$ templates, we remove the substring predicates and predicates involving multiple columns, 
e.g., \textsf{inv\_quantity\_on\_hand} < \textsf{cs\_quantit}. We generate $100$ queries for each template by the provided query generation tool.  
We only preserve unique queries with nonzero cardinality. 
Table~\ref{tbl:querysets} summarizes the 4 corresponding query sets. 

\stitle{Implementation details.}
\CardOOD is built on \PyTorch~\cite{pytorch} and \Pandas.
We use the Adam~\cite{DBLP:journals/corr/KingmaB14} optimizer with a decaying learning rate to train our models, where training hyper-parameters are turned in their empirical ranges: initial learning rate  $\in [10^{-3}, 0.5\times 10^{-3}, 10^{-4}]$, batch size $\in [32, 64, 128]$, epochs $\in [80, 100, 120, 150]$. We set the L2 penalty of Adam to $10^{-4}$ and the decaying factor to $0.85$. For the configurations of the MSCN and MLP models, the hidden dimensions of the set convolutions and MLP layers are set to 64 and 128, respectively.
As \dsb still contains complicated predicate filters after preprocessing,
we use the source code from \cite{DBLP:journals/pvldb/NegiWKTMMKA23}
to encode the queries and build MSCN models. Following \cite{DBLP:journals/pvldb/NegiWKTMMKA23}, estimated cardinalities 
by \PostgreSQL are injected into the input query features to enhance the representation.
The best models are chosen by grid search. 
The parameters of different algorithms in \CardOOD are referred to their original papers. 
Here, we set the weight $\lambda$ of the CORAL (Eq.~(\ref{eq:loss:align})) and the order loss (Eq.~(\ref{eq:loss:order})) terms as $0.5$. 
For DANN, the weight of the CE loss (Eq.~(\ref{eq:loss:cla})) is set to $10^{-2}$ for MLP and $10^{-3}$ for MSCN, respectively.
For Query Mixup, the hyper-parameter $\alpha$ in the Beta distribution is set to 0.5 and the Gaussian bandwidth $\sigma$ is set to 0.1.
For OrderEmb, we set the number of contrastive queries to $5$ by default for each training query.
By default, both model training and prediction are conducted on a Tesla V100 with 16GB GPU memory.
The experiments for query optimization in an RDBMS are conducted on a Linux server with 32 Intel(R) E5-2620 v4 CPUs and 256GB RAM, running \PostgreSQL 12.4. We use the default configuration of \PostgreSQL so that the estimation statistics are updated during query execution, which may improve the estimation of the built-in estimator. For each query, we clean up the shared memory of \PostgreSQL and the Linux kernel buffer to fulfill cold-start. 
We do not build indexes for base tables as incorporating the indexes introduces extra search space and makes the optimization more complex \cite{DBLP:conf/sigmod/LiLK22}. 

\begin{table}[t]
{\footnotesize
\caption{Query Sets}
\label{tbl:querysets}
\begin{center}
    \begin{tabular}{c c c c c}
    \toprule
     {\bf Type} & {\bf Dataset}  & {\bf \# Queries} & \makecell{ \bf \# Sel. / Join / Templ.}  & \makecell{ \bf Range of \bf $c(q)$} 
    \\ \midrule                                             
   Single Rel. & \forest  & 18,000 & \{2, $\cdots$, 10\} & $[10^0, 10^6]$  \\  
   Join & \imdb  & 15,000 & \{0, 1, 2, 3, 4\} & $[10^{0}, 10^{8}]$  \\
   Join & \job  & 84,000 & \{t1, t2, $\cdots$, t70\} & $[ 10^{0}, 10^{8} ]$ \\
   Join & \dsb & 13,00 & \{t1, t2, $\cdots$, t13\}  & $[10^{0}, 10^{8} ]$ \\
   \bottomrule
   \end{tabular}
\end{center}
}
\end{table}

\subsection{Prediction Robustness}
\label{sec:exp:robust}
We investigate the prediction robustness of \CardOOD under shift training query workloads and query templates.
To control the discrepancy of distributions, we distinguish simple/complex queries from the  query sets regarding the numbers of selection/join conditions.
For single relation \forest, the queries with $\{2,\cdots 6\}$ selection conditions are regarded as simple queries, while queries with $\{7,\cdots 10\}$ selection conditions are complex queries. 
For multi-relation join queries of \imdb,
the queries with $\{0, 1, 2\}$ join conditions are regarded as simple queries, while queries with $\{3, 4\}$ join conditions are complex queries.
We preserve uniform distributed $10\%$ queries as the test queries, and construct training workloads by varying the ratio of simple/complex queries in $\{$20\%/80\%, 50\%/50\%, 80\%/20\%$ \}$, respectively.
  
\comment{
we introduce skew into the training query set. Specifically, we classify queries into simple queries and complex queries. Queries with a selection condition number larger than $\floor{(2 + D) / 2}$ or a join condition number larger than $\floor{(0 + 4) / 2}$ are categorized as complex queries. All other queries are categorized as simple queries.

We then apply skew to the training query set as follows: for a given skew ratio $s \in [0, 1]$, we randomly select $s \cdot N$ queries from simple query set and $(1 - s) \cdot N$ queries from complex query set, where $N$ is the total number of queries with a specific selection or join condition number. This method allows us to create training query set with varying skew ratios. 
We consider the following OOD scenarios based on skewed query set:
}

\stitle{Distribution on the \# selection condition is OOD.} 
We evaluate the models trained by different algorithms on \forest with different ratios of simple/complex queries. 
Table \ref{tab:qerr_MLP_forest} and Table \ref{tab:qerr_MSCN_forest}
show the prediction \Qerror of MLP and MSCN on \forest, respectively. 
Here, the \Qerror of the six algorithms is highlighted when it is significantly lower than that of the ERM baseline ($20\%$ lower). 
In general, there is no single approach that consistently outperforms others in all cases. However, we can 
acquire some findings from the results. 
When using MLP as the base model (Table~\ref{tab:qerr_MLP_forest}), all the six learning algorithms manifest superior performance over ERM at the $95\%$ and $95\%$ quantiles, validating 
the effectiveness of the learning strategies tailored for OOD scenarios. 
Among these algorithms, OrderEmb is able to retain considerately well performance even in the worst-case scenarios. 
On the contrary, ERM tends to perform worse on simple 
queries for a skew ratio of $20\%/80\%$ and $80\%/20\%$, 
which indicates that a plain learning strategy without specific techniques for OOD is heavily affected by the imbalanced training query workload.
When MSCN serves as the base model (Table~\ref{tab:qerr_MSCN_forest}), generally all the learning algorithms, including ERM, enjoy a performance gain compared with MLP. 
That is because MSCN is originally designed for SQL query cardinality estimation and possesses more powerful modeling capabilities. 
In most cases, the learning strategies considering OOD can still outperform ERM, while the improvement
is not as remarkable as that by MLP. 
The reason for the marginal improvements would lie in the effectiveness of the MSCN model, which limits the potential for robust learning algorithms to make significant contributions.

\stitle{Distribution on the \# join condition is OOD.} We evaluate 
the models trained by different algorithms on \imdb queries with different ratios of the simple/complex queries.
Table \ref{tab:qerr_MLP_IMDB} and Table \ref{tab:qerr_MSCN_IMDB} 
show the \Qerror for MLP and MSCN, respectively. 
Overall, the six algorithms outperform ERM significantly 
in most cases of large quantiles, for both MLP and MSCN models. 
The result is different from that 
in the single table queries, as multi-relation join queries are more complicated than single relation queries, highlighting the efficacy of the new learning strategies for alleviating the OOD problem. 
We observe that the accuracy of the algorithms in \CardOOD is consistently worse on complex queries 
than that on simple queries at the $75\%$, $95\%$ and $99\%$ quantiles, regardless of the degree of distribution shift, with only one exception, i.e., Group DRO with the ratio of $20\%/80\%$ at the $75\%$ quantile.  
That further reflects the predictions for multi-join queries become more challenging 
as the number of join conditions increases.
In our testing, Query Masking suffers from the risk of underestimation due to the dropout training mechanism.
Our proposed algorithm, OrderEmb, performs well compared to other
learning algorithms, although it is not the best in all cases. Similar to the results on \forest, OrderEmb can achieve reasonably well performance on both simple and complex queries. 

\stitle{Test Query Template is OOD}. 
We evaluate the models trained by different algorithms on \dsb.
We randomly select $3$ query templates with $5$ 
tables for testing and use the remaining $10$ templates for training.
Table \ref{tab:qerr_MSCN_DSB} shows the \Qerror for MSCN on the test queries.
Overall, the six learning strategies can outperform ERM in most cases. 
Considering that \dsb contains complex hand-written query templates, the plain ERM strategy is insufficient for handling distribution shifts in query templates, especially for template 040.
We observe that Group DRO and Query Mixup perform poorly at the
median but achieve better results at the $99\%$ quantile.
Although they perform worse than ERM at some quantiles, the smaller range of errors achieves robustness over the testing queries. In contrast, DANN and OrderEmb achieve the best 
performance on $50\%$ quantile, while exhibiting larger error ranges. When delving into individual query templates, we find  that the performance gain mainly comes from template $040$. 
The reason would be that template $040$ contains a
complicated predicate involving date comparison, 
which is unseen in the training set.

\stitle{Takeaway.} The learning strategies in \CardOOD outperform ERM in most cases of $95\%$ and $99\%$ quantiles of the predictions, even in the non-OOD cases. Generally, 
Deep CORAL and OrderEmb achieve the overall best performance in single relation queries, while Query Mixup and OrderEmb achieve the best performance for multi-join queries.
In practice, the performance gain of \CardOOD is also influenced by the query type, the model type and the degree of distribution shift. 

\begin{table*}[t]
\centering
\small
\caption{Q-error for MLP on \forest}\label{tab:qerr_MLP_forest}
\begin{tabular}{cc|rrrr|rrrr|rrrr}
\toprule
 \multirow{2}{*}{\makecell[c]{Ratio of \\ Simple/Complex}} & \multirow{2}{*}{Algorithm} & \multicolumn{4}{c|}{Overall} & \multicolumn{4}{c|}{Simple Queries} & \multicolumn{4}{c}{Complex Queries} \\

 &       & {50\%} & {75\%} & {95\%} & {99\%} & {50\%} & {75\%} & {95\%} & {99\%} & {50\%} & {75\%} & {95\%} & {99\%} \\\midrule

\multirow{7}{*}{$20\%/80\%$} &  ERM  & 1.40  & 7.43  & 102.48  & 976.64  & 1.78  & 9.81  & 149.23 & 2233.92  & 1.04 & 4.68  & 51.13 & 499.32  \\ \cline{2-14}
    & Deep CORAL  & 1.46 & 6.20 & \cellcolor{LightGreen} 43.16 & \cellcolor{LightGreen} 200.54   & 1.79 & \cellcolor{LightGreen} 6.72 & \cellcolor{LightGreen} 51.12 & \cellcolor{LightGreen} 236.02  & 1.20 & 5.21 & \cellcolor{LightGreen} 34.33 & \cellcolor{LightGreen} 89.58           \\
    & DANN &  1.38 & \cellcolor{LightGreen}3.43 & \cellcolor{LightGreen}81.50 & 860.27  & 1.99 & \cellcolor{LightGreen}2.37 & \cellcolor{LightGreen}103.71 & \cellcolor{LightGreen} 1059.92 & 1.11 & 4.81  & 61.27  & \cellcolor{LightGreen} 363.87                      \\
    & Group DRO & \cellcolor{LightGreen} 1.06 & \cellcolor{LightGreen} 4.58 & \cellcolor{LightGreen} 39.57 & \cellcolor{LightGreen} 351.06 & \cellcolor{LightGreen} 1.09 & \cellcolor{LightGreen} 4.87 & \cellcolor{LightGreen} 48.21 & \cellcolor{LightGreen} 408.27 & 1.24 & 5.06 & \cellcolor{LightGreen} 34.42 & \cellcolor{LightGreen} 144.46                    \\
    & Query Mixup & 1.30  & \cellcolor{LightGreen} 4.41 & \cellcolor{LightGreen} 43.55 & \cellcolor{LightGreen} 230.09  & \cellcolor{LightGreen} 1.42 & \cellcolor{LightGreen} 4.79 & \cellcolor{LightGreen} 61.62 & \cellcolor{LightGreen} 268.97 & 1.16 & 3.83 & \cellcolor{LightGreen} 25.34 & \cellcolor{LightGreen} 97.34                  \\ 
     & Query Mask  & 1.39 & 7.36 & 83.52 & \cellcolor{LightGreen} 676.26 &  1.97 & \cellcolor{LightGreen} 9.74 & 132.65 & \cellcolor{LightGreen} 1164.76 & 1.01 & 4.32 & 42.78 & \cellcolor{LightGreen} 336.94                     \\ 
    & OrderEmb  & 1.19 & \cellcolor{LightGreen} 4.72 & \cellcolor{LightGreen} 40.62 & \cellcolor{LightGreen} 227.28 & \cellcolor{LightGreen} 1.09 & \cellcolor{LightGreen} 4.84 & \cellcolor{LightGreen} 77.37 & \cellcolor{LightGreen} 491.61  & 1.36 & 5.41 & \cellcolor{LightGreen} 31.14 & \cellcolor{LightGreen} 138.06  \\
\midrule
\multirow{7}{*}{$50\%/50\%$} &  ERM  & 1.20 & 6.64  & 109.60 & 1177.61 & 1.10  & 6.00  & 125.55  & 1922.73  & 1.43  & 7.18  & 83.32 & 641.37   \\ \cline{2-14}
    & Deep CORAL  & 1.04  & 6.19 & \cellcolor{LightGreen} 68.57 & \cellcolor{LightGreen} 343.64 & 1.63 & 7.36 & \cellcolor{LightGreen} 72.91 & \cellcolor{LightGreen} 358.41 & 2.07  & 10.23 & \cellcolor{LightGreen} 61.68  & \cellcolor{LightGreen} 234.91      \\
    & DANN & 1.42 & 5.23 & \cellcolor{LightGreen} 52.51 &  \cellcolor{LightGreen} 262.52  & 2.11  & 7.52 & \cellcolor{LightGreen} 66.90 & \cellcolor{LightGreen} 262.52 & 1.35 & 3.03  & \cellcolor{LightGreen} 28.09  & \cellcolor{LightGreen} 211.52                \\
    & Group DRO & 1.07 & \cellcolor{LightGreen} 4.86 & \cellcolor{LightGreen} 63.08 & \cellcolor{LightGreen} 614.49 & 1.07 & \cellcolor{LightGreen} 4.80 & \cellcolor{LightGreen} 68.44 & \cellcolor{LightGreen} 1112.92 & 1.58 & 6.39 & 66.92 &  \cellcolor{LightGreen} 467.59                 \\
    & Query Mixup   & 1.13  & \cellcolor{LightGreen} 4.65  & \cellcolor{LightGreen} 45.12  & \cellcolor{LightGreen} 266.13  & 1.02 & \cellcolor{LightGreen} 4.43 & \cellcolor{LightGreen} 69.38 & \cellcolor{LightGreen} 310.50 & 1.33 & \cellcolor{LightGreen} 5.04 & \cellcolor{LightGreen} 32.69 & \cellcolor{LightGreen} 146.58                     \\ 
    & Query Mask  & 1.63  & 7.02 & \cellcolor{LightGreen} 68.22 & \cellcolor{LightGreen} 360.43 & 2.69  & 9.13  & \cellcolor{LightGreen} 83.58 & \cellcolor{LightGreen} 358.27 & 1.24 & 6.43 & \cellcolor{LightGreen} 56.07 & \cellcolor{LightGreen} 360.78                      \\
    & OrderEmb   & 1.29 & \cellcolor{LightGreen} 5.27 & \cellcolor{LightGreen} 37.60 &  \cellcolor{LightGreen} 240.89  & 1.24 & 4.96 & \cellcolor{LightGreen} 69.51  & \cellcolor{LightGreen} 357.69  & 1.30  & \cellcolor{LightGreen} 5.61  & \cellcolor{LightGreen} 32.33  & \cellcolor{LightGreen} 103.51  \\
\midrule
\multirow{7}{*}{$80\%/20\%$} &  ERM  &  1.24 & 5.82  & 86.02 & 1099.84 & 1.29 & 5.04 & 70.00 & 568.80  & 1.18  & 7.19  & 127.13  & 1772.14  \\ \cline{2-14}
     & Deep CORAL  & 1.22  & 4.68 & \cellcolor{LightGreen} 36.58 & \cellcolor{LightGreen} 253.84 & 1.19 & 4.47 & \cellcolor{LightGreen} 48.18 & \cellcolor{LightGreen} 368.54 & 1.23 & \cellcolor{LightGreen} 4.85 & \cellcolor{LightGreen} 40.87  & \cellcolor{LightGreen} 140.59            \\
    & DANN &  1.27 & \cellcolor{LightGreen}2.87 &\cellcolor{LightGreen} 24.36 & \cellcolor{LightGreen}171.40 & \cellcolor{LightGreen}1.03  & \cellcolor{LightGreen}3.46 & \cellcolor{LightGreen}29.01 & \cellcolor{LightGreen}232.65 & 1.92 & \cellcolor{LightGreen}2.35 & \cellcolor{LightGreen} 20.42 & \cellcolor{LightGreen}98.19       \\
    
    & Group DRO & 1.08 & \cellcolor{LightGreen} 4.43 & \cellcolor{LightGreen} 45.82 & \cellcolor{LightGreen} 199.41 &  1.15 & \cellcolor{LightGreen} 3.36 & \cellcolor{LightGreen} 38.57 & \cellcolor{LightGreen} 200.27 & 1.48 & 6.82 & \cellcolor{LightGreen} 54.62 & \cellcolor{LightGreen} 183.04                  \\
    & Query Mixup & 2.13 & 8.65 & 104.76 & \cellcolor{LightGreen} 487.68 & 1.98 & 7.68 & 104.81 & 509.59 & 2.46 & 9.52 & \cellcolor{LightGreen} 101.59 & \cellcolor{LightGreen} 481.96          \\ 
    
    & Query Mask  & 1.50  & 6.82 & 90.84 & 1062.46 &  1.63 & 6.50 & 82.89 & 577.49 & 1.35 & 7.59 & 107.66 & \cellcolor{LightGreen} 1217.31                     \\ 
    & OrderEmb  & 1.04 & \cellcolor{LightGreen} 3.80  & \cellcolor{LightGreen} 29.79 & \cellcolor{LightGreen} 192.90   & 1.43 & \cellcolor{LightGreen} 3.85 & \cellcolor{LightGreen} 28.44 & \cellcolor{LightGreen} 233.16 & 1.32 & \cellcolor{LightGreen} 5.51 & \cellcolor{LightGreen} 31.54 & \cellcolor{LightGreen} 190.85 \\
\bottomrule

\end{tabular}
\end{table*}

\begin{table*}[t]
\centering
\small
\caption{Q-error for MSCN on \forest}\label{tab:qerr_MSCN_forest}
\begin{tabular}{cc|rrrr|rrrr|rrrr}
\toprule
 \multirow{2}{*}{\makecell[c]{Ratio of \\ Simple/Complex}} & \multirow{2}{*}{Algorithm} & \multicolumn{4}{c|}{Overall} & \multicolumn{4}{c|}{Simple Queries} & \multicolumn{4}{c}{Complex Queries} \\

  &     & {50\%} & {75\%} & {95\%} & {99\%} & {50\%} & {75\%} & {95\%} & {99\%} & {50\%} & {75\%} & {95\%} & {99\%} \\\midrule

\multirow{7}{*}{$20\%/80\%$} &  ERM  &  1.30 & 4.64 & 24.56 & 110.06 & 1.32 & 4.26 & 31.39 & 225.45 & 1.27  & 5.63 & 29.97 & 110.03  \\ \cline{2-14}
    & Deep CORAL  & 1.39  &  \cellcolor{LightGreen} 2.44 & \cellcolor{LightGreen} 16.75 & \cellcolor{LightGreen} 78.25 & 2.04  & \cellcolor{LightGreen} 1.85 & \cellcolor{LightGreen} 14.93 & \cellcolor{LightGreen} 95.80  & 1.05  & \cellcolor{LightGreen} 3.21  & \cellcolor{LightGreen} 17.75  & \cellcolor{LightGreen} 60.89           \\
    & DANN & \cellcolor{LightGreen}  1.04 & 4.51 & 29.79 & 122.50 &  1.17 &  4.10 & 39.09 & 277.94 & 1.15 & 5.12 & 25.02 &  \cellcolor{LightGreen} 86.74                  \\
    & Group DRO & 1.05 & 4.45 & 38.78 & 170.87 & 1.15 & 4.04 & 47.20 & 380.66 &  1.11 & 5.05 & 26.50 & 119.99                      \\
    & Query Mixup & 1.36 & 6.16 & 49.33 & 318.92 &  1.36 & 6.12 & 68.72 & 412.20 & 1.36 & 6.12 & 68.72 &  412.20                    \\
    & Query Mask & 1.41 &  5.02 & 28.12 &  102.00 & 1.42 & 4.54 & 29.47 & 210.30  & 1.41 & 6.65 & 33.92 & 130.55     \\ 
    & OrderEmb & 1.12 & 4.15 & 33.68 & 162.00 & 1.17 & 3.93 & 39.89 & 271.80 & 1.02 & 4.72 & 24.12 & 92.34 \\
\midrule
\multirow{7}{*}{$50\%/50\%$} &  ERM  & 1.18 & 5.75 & 41.40 & 182.31 & 1.02 & 4.46 & 51.04 & 320.56  & 1.54 & 7.68 & 37.45  & 133.24 \\ \cline{2-14}
    & Deep CORAL  & 1.02  & \cellcolor{LightGreen} 3.16  & \cellcolor{LightGreen} 20.81  & \cellcolor{LightGreen} 85.77   & 1.75  & \cellcolor{LightGreen} 1.89  & \cellcolor{LightGreen} 13.38 & \cellcolor{LightGreen} 77.84  & 1.69  & \cellcolor{LightGreen} 4.69  & \cellcolor{LightGreen} 28.67  &  \cellcolor{LightGreen} 91.49          \\
    & DANN & 1.14 & 5.35 & 41.02  & 166.70 & 1.01 & 4.45 & 46.02 & 301.44 & 1.44 & 6.85 & 35.68 & 111.43                     \\
    & Group DRO & 1.18 & 5.70 & 45.08 & 197.07 & 1.13 & 3.97 & 54.20 & 377.81 &  1.69 & 8.15 & 39.73 &  148.60                    \\
    & Query Mixup  & 1.40 & 6.60  & 48.09 & 245.58 &  1.26 & 5.69 & 58.05 & 418.24 & 1.68 & 7.54 & 38.98 & 110.34                  \\
    & Query Mask   & 1.12  & 5.22 & 38.65 &  159.19 & 1.01 & 4.27 & 47.44 & 316.59 & 1.39 & 6.46 & 31.47 & 114.46               \\ 
    & OrderEmb  & 1.18 & 5.02 & 37.96 & 154.20 & 1.00 & 4.26 & 42.82 & 281.15 & 1.52 & 6.33 & 32.53 & \cellcolor{LightGreen} 96.02 \\
\midrule
\multirow{7}{*}{$80\%/20\%$} &  ERM  & 1.11 & 4.24 & 28.67 & 127.21 & 1.14 & 3.92 & 37.13 & 272.20  & 1.04 & 4.88 & 28.52 & 98.17  \\ \cline{2-14}
    & Deep CORAL  &  1.10 & \cellcolor{LightGreen} 2.72  & \cellcolor{LightGreen} 18.36  &  \cellcolor{LightGreen} 74.91  & 1.51  & \cellcolor{LightGreen} 1.67  & \cellcolor{LightGreen} 12.02 & \cellcolor{LightGreen} 60.05  & 1.46  & 4.35  & 25.07  & 82.86           \\
    & DANN & 1.04  & 4.51 & 29.79 & 122.50 & 1.17 & 4.10 & 39.09 & 277.94 & 1.15 & 5.12 & 25.02 & 86.74                     \\
    & Group DRO & 1.07 & 4.99 & 42.03 & 184.25 & 1.07 & 4.18 & 50.60 & 399.44 &  1.32 & 6.02 & 30.28 & 134.62                     \\
    & Query Mixup  & 1.52 & 7.08 & 55.50 & 282.83 & 1.46 & 6.52 & 76.19 & 443.81 &  1.73 & 7.55 & 41.89 & 129.80                      \\
    & Query Mask & 1.15 & 4.13 & 27.55 &  121.48 & 1.18 &  3.79 & 36.05 & 271.93 & 1.08 & 5.22 & 26.45 & 103.16                     \\
    & OrderEmb  & 1.15 & 4.08  & 29.07 &  126.76 & 1.21 & 3.87 & 35.42 & 269.81 & 1.10 & 4.75 & 23.81 & 97.64 \\
\bottomrule

\end{tabular}
\end{table*}

\begin{table*}[t]
\centering
\small
\caption{Q-error for MLP on \imdb}\label{tab:qerr_MLP_IMDB}

\begin{tabular}{cc|rrrr|rrrr|rrrr}
\toprule
\multirow{2}{*}{\makecell[c]{Ratio of \\ Simple/Complex}} & \multirow{2}{*}{Algorithm} & \multicolumn{4}{c|}{Overall} & \multicolumn{4}{c|}{Simple Queries} & \multicolumn{4}{c}{Complex Queries} \\

 &   & {50\%} & {75\%} & {95\%} & {99\%} & {50\%} & {75\%} & {95\%} & {99\%} & {50\%} & {75\%} & {95\%} & {99\%} \\\midrule

\multirow{7}{*}{$20\%/80\%$} &  ERM  & 1.04  & 2.22  & 21.83 & 161.88  & 1.11 & 1.71 & 5.39 & 23.56 & 1.41  & 5.24 & 76.47 & 450.43  \\ \cline{2-14}
    & Deep CORAL  & 1.27 & 2.99 & 17.54 & \cellcolor{LightGreen} 104.84 & 1.25 & 2.33 & 8.03 & 42.01 & 1.29 & 5.21 & \cellcolor{LightGreen} 31.81 & \cellcolor{LightGreen} 344.91        \\
    & DANN & 1.18 & 2.27  & \cellcolor{LightGreen} 15.62 & \cellcolor{LightGreen} 111.52  & 1.17 & 1.86 & 6.50 & 28.28 & 1.23 & \cellcolor{LightGreen} 3.71 & \cellcolor{LightGreen} 41.82 & 435.90                     \\
    & Group DRO & 1.37 & 3.51 & 27.54 &  233.95  & 1.47 &  3.69 & 33.35 & 88.01 & \cellcolor{LightGreen} 1.03 & 4.51  &  68.22 & 572.06                     \\
    & Query Mixup & 1.29 & 2.25 & \cellcolor{LightGreen} 11.97 & \cellcolor{LightGreen} 75.28 & 1.37 & 2.05 & 7.31 & 37.89 & 1.12 & \cellcolor{LightGreen} 2.99 & \cellcolor{LightGreen} 19.52 & \cellcolor{LightGreen} 117.82                    \\
    &Query Mask & 1.15 & 2.22 & \cellcolor{LightGreen} 17.35 & 171.04 & 1.17 & 1.90 & 5.83 & 26.15 & \cellcolor{LightGreen} 1.07 & \cellcolor{LightGreen} 3.87 & \cellcolor{LightGreen} 41.28 & \cellcolor{LightGreen} 353.90                      \\
    & OrderEmb &  1.68 & 3.10  & \cellcolor{LightGreen} 11.05 & \cellcolor{LightGreen} 100.93 & 1.79  & 2.95 & 5.88 & \cellcolor{LightGreen} 16.77  & 1.36 & \cellcolor{LightGreen} 3.93 & \cellcolor{LightGreen}  41.51  & \cellcolor{LightGreen} 239.55 \\
\midrule
\multirow{7}{*}{$50\%/50\%$} &  ERM  & 1.08  & 2.21  & 22.81  & 217.30   & 1.08 & 1.71 & 6.98 & 30.85 & 1.10 & 4.21  & 92.21  & 438.39  \\ \cline{2-14}
    & Deep CORAL & 1.19 & 2.22 & \cellcolor{LightGreen} 14.24 & \cellcolor{LightGreen} 119.25  & 1.15 & 1.72 & 7.00 &  28.58 &  1.46 & 3.93 & \cellcolor{LightGreen} 41.92 & \cellcolor{LightGreen} 329.07           \\
    & DANN & 1.23 & 2.55 & \cellcolor{LightGreen} 13.67 & \cellcolor{LightGreen} 107.06 & 1.15 & 1.83 & \cellcolor{LightGreen} 5.29 & 31.72 & 1.59 & 5.42  & \cellcolor{LightGreen} 32.71 & 402.91                  \\
    & Group DRO & 1.36 & 3.86 & 26.25 & 236.52 & 1.40 & 3.29 & 22.75 & 60.95 & 1.18 & 4.71 & 78.76 & 446.71                     \\
    & Query Mixup & 1.17 & 1.85 & \cellcolor{LightGreen} 15.12 & \cellcolor{LightGreen} 113.09 & 1.25 & 1.73 & \cellcolor{LightGreen} 4.60 & \cellcolor{LightGreen} 21.37 & 1.07 &  3.71 & \cellcolor{LightGreen} 31.97 & \cellcolor{LightGreen} 191.33                     \\
    & Query Mask & 1.42  & 2.74 & \cellcolor{LightGreen} 16.90 & \cellcolor{LightGreen} 115.54 & 1.37 & 2.18 & 5.12 & \cellcolor{LightGreen} 20.46 & 1.82 & 6.57  & \cellcolor{LightGreen} 36.94 & \cellcolor{LightGreen} 326.10                     \\
    & OrderEmb    & 1.06 & 1.97 & 20.41 & \cellcolor{LightGreen} 130.12 & 1.01 & 1.50  & 5.84 & 32.76 & 1.32 & 4.30 & \cellcolor{LightGreen} 48.29 & \cellcolor{LightGreen} 303.65  \\
\midrule
\multirow{7}{*}{$80\%/20\%$} &  ERM  & 1.20 & 2.16 & 24.75 &  441.50  & 1.17  & 1.62 & 4.93 & 21.07  & 1.47  & 6.60  & 129.44 & 3259.70  \\ \cline{2-14}
    & Deep CORAL  &  1.24 & 2.58 & \cellcolor{LightGreen} 17.60 & \cellcolor{LightGreen} 118.13  & 1.14 & 1.78 & 6.07 & 34.76 & 1.85 & 6.68 & \cellcolor{LightGreen} 61.22 & \cellcolor{LightGreen} 426.06          \\
    & DANN & 1.18 & 2.37 & \cellcolor{LightGreen} 15.65 & \cellcolor{LightGreen} 173.28 & 1.13 & 1.83 & 5.05 & 26.51 & 1.39 & 5.68  & \cellcolor{LightGreen} 63.42 & \cellcolor{LightGreen} 669.89                    \\
    & Group DRO & 1.24 & 3.46 & 35.54 &  714.88  &  1.24 & 2.91 & 14.43 & 56.11 & 1.31 & 6.21  & 117.44 & \cellcolor{LightGreen} 1883.77                     \\
    & Query Mixup & 1.21 & 2.39 & \cellcolor{LightGreen} 15.55 & \cellcolor{LightGreen} 83.87 & 1.06  & 1.55 & 5.67 & 23.20 & 2.05 & 5.40 & \cellcolor{LightGreen} 32.28  & \cellcolor{LightGreen} 262.46                    \\
    & Query Mask  & 1.29 & 2.19 & 40.03 & 987.60 & 1.36 & 1.99 & 4.63 & 22.79 & 1.22 & 6.50 & 149.96 &  6337.39                    \\
    & OrderEmb  & 1.01  & 2.30 & 22.24 & \cellcolor{LightGreen}  131.93 & 1.00  & 1.80 & 9.75 & 68.55 & \cellcolor{LightGreen} 1.06 & \cellcolor{LightGreen} 4.12 & \cellcolor{LightGreen} 55.06 & \cellcolor{LightGreen} 331.37  \\
\bottomrule

\end{tabular}
\end{table*}

\begin{table*}[t]
\centering
\small
\caption{Q-error for MSCN on \imdb}\label{tab:qerr_MSCN_IMDB}

\begin{tabular}{cc|rrrr|rrrr|rrrr}
\toprule
 \multirow{2}{*}{\makecell[c]{Ratio of \\ Simple/Complex}} & \multirow{2}{*}{Algorithm} & \multicolumn{4}{c|}{Overall} & \multicolumn{4}{c|}{Simple Queries} & \multicolumn{4}{c}{Complex Queries} \\

  &      & {50\%} & {75\%} & {95\%} & {99\%} & {50\%} & {75\%} & {95\%} & {99\%} & {50\%} & {75\%} & {95\%} & {99\%} \\\midrule

\multirow{7}{*}{$20\%/80\%$} &  ERM  & 1.08 & 2.46 & 25.55 &  241.01 & 1.20 & 2.10 & 11.29 & 70.23 & 1.15 & 4.55 & 66.69 & 1016.23 \\ \cline{2-14}
    & Deep CORAL  &  1.02 & \cellcolor{LightGreen} 1.89 & \cellcolor{LightGreen}10.91  & \cellcolor{LightGreen} 84.1  & 1.01  & \cellcolor{LightGreen}1.58  & \cellcolor{LightGreen}4.76  & \cellcolor{LightGreen}22.71  & 1.02  & \cellcolor{LightGreen} 2.28  & \cellcolor{LightGreen}16.02  & \cellcolor{LightGreen} 117.02          \\
    & DANN & 1.36 & 2.95 & \cellcolor{LightGreen} 10.97 & \cellcolor{LightGreen} 66.99 & 1.14 &  2.07 & \cellcolor{LightGreen} 5.51 & \cellcolor{LightGreen} 18.03  & 2.09 & 5.43 & \cellcolor{LightGreen} 22.60 & \cellcolor{LightGreen} 276.05                     \\
    & Group DRO & 1.17 & 3.81 & 24.77 &   205.05 & 1.55 & 4.55 & 17.19 & 82.32 & 1.14 & 4.32 & 53.95 & \cellcolor{LightGreen} 269.72                     \\
    & Query Mixup & 1.01  & 2.06 & \cellcolor{LightGreen} 9.63 & \cellcolor{LightGreen} 61.56 & 1.14 & 2.12 & \cellcolor{LightGreen} 6.23 &  \cellcolor{LightGreen} 17.09 & 1.23 & \cellcolor{LightGreen} 2.82 & \cellcolor{LightGreen} 21.04  & \cellcolor{LightGreen} 141.37                    \\
    & Query Mask & 1.11 & 2.74 & 24.92 & 282.36 & 1.29 & 2.28 & 13.08 & 72.34 & 1.23 & 4.11 & 67.99 & 1001.82 \\
    & OrderEmb  & 1.02 & 2.12 & \cellcolor{LightGreen} 9.53 & \cellcolor{LightGreen} 64.40 & 1.11 & 2.06 & \cellcolor{LightGreen} 6.93 & \cellcolor{LightGreen} 14.23 & 1.13 & \cellcolor{LightGreen} 2.38 & \cellcolor{LightGreen} 27.95 & \cellcolor{LightGreen} 147.62 \\
\midrule
\multirow{7}{*}{$50\%/50\%$} & ERM & 1.05 & 2.33 &  20.18 &  196.03 & 1.06 & 2.18 & 9.94 & 61.56 & 1.58 & 4.89 & 54.62 &  946.64 \\ \cline{2-14}
    & Deep CORAL  &  1.09 & \cellcolor{LightGreen}1.55  & \cellcolor{LightGreen}7.46  & \cellcolor{LightGreen}39.79   &  1.13 & \cellcolor{LightGreen} 1.32 & \cellcolor{LightGreen} 3.18 & \cellcolor{LightGreen}12.86  & \cellcolor{LightGreen}1.05  & \cellcolor{LightGreen}1.95  & \cellcolor{LightGreen} 10.16  & \cellcolor{LightGreen} 55.21           \\
    & DANN & 1.14 & 2.70 & 19.50 & \cellcolor{LightGreen} 148.07 & 1.25 & 2.55 & 9.78 & \cellcolor{LightGreen} 43.46  & \cellcolor{LightGreen} 1.17 & \cellcolor{LightGreen} 3.42  & 48.46  & \cellcolor{LightGreen} 459.45                    \\
    & Group DRO & 1.02 & 3.21 & 34.33 &  211.07 & 1.33 & 3.69 & 15.74 & 49.26 & 1.57 & 5.57 & 85.64 & \cellcolor{LightGreen}  495.89                    \\
    & Query Mixup & 1.11 & 2.23 & \cellcolor{LightGreen} 11.60 & \cellcolor{LightGreen} 81.95 & 1.28 & \cellcolor{LightGreen} 2.18 & \cellcolor{LightGreen} 6.18 & \cellcolor{LightGreen} 21.83 & \cellcolor{LightGreen} 1.23 & \cellcolor{LightGreen} 2.66 & \cellcolor{LightGreen} 30.11 & \cellcolor{LightGreen} 123.27                   \\
    & Query Mask & 1.05 & 2.46 & 22.97 & 191.84 & 1.15 & 2.32 & 11.35 & 67.10 & 1.38 & 4.54 & 59.85 & 1089.74                     \\
    & OrderEmb & 1.13 & 2.08 & \cellcolor{LightGreen} 8.20 & \cellcolor{LightGreen} 71.77 & 1.02 & \cellcolor{LightGreen}  1.73 & \cellcolor{LightGreen} 5.81 & \cellcolor{LightGreen} 12.89 & 1.45 & \cellcolor{LightGreen} 3.04 & \cellcolor{LightGreen} 21.32 & \cellcolor{LightGreen} 134.54 \\
\midrule
\multirow{7}{*}{$80\%/20\%$} &  ERM  & 1.17 & 2.76 & 27.74 &  308.11 &  1.26 & 2.45 & 11.91 & 81.89 & 1.17 & 5.08 &  82.07 & 1372.75 \\ \cline{2-14}
    & Deep CORAL  & 1.01  & 2.53 & 26.58  & 265.04   & 1.16  & \cellcolor{LightGreen} 1.50  & \cellcolor{LightGreen}6.69 & \cellcolor{LightGreen} 35.68  & 1.19  & \cellcolor{LightGreen}3.71  & \cellcolor{LightGreen}50.26  & \cellcolor{LightGreen} 564.97           \\
    & DANN & 1.21 & 2.74 & \cellcolor{LightGreen} 22.04 & \cellcolor{LightGreen} 188.10 & 1.39 & 2.66 & 9.90 & \cellcolor{LightGreen} 47.58 & 1.10 & \cellcolor{LightGreen} 3.93 & \cellcolor{LightGreen} 56.23 & \cellcolor{LightGreen} 429.83                     \\
    & Group DRO & 1.34 & 3.18 & 27.54 & \cellcolor{LightGreen}  228.10 & 1.64 & 3.47 & 9.80 & \cellcolor{LightGreen} 57.26 & 1.25 & 4.79 & 78.21 & \cellcolor{LightGreen} 741.29                     \\
    & Query Mixup & 1.16 & \cellcolor{LightGreen} 2.14 & \cellcolor{LightGreen} 13.55 & \cellcolor{LightGreen} 113.63 & 1.23 & 2.10 & \cellcolor{LightGreen} 6.14 & \cellcolor{LightGreen} 17.56  & 1.04 & \cellcolor{LightGreen} 3.01 & \cellcolor{LightGreen} 33.43 &  \cellcolor{LightGreen} 334.33                     \\
    & Query Mask & 1.10 & 2.63 & 27.61  &  279.09 & 1.24 & 2.37 & 11.32 & \cellcolor{LightGreen} 64.92 & 1.32 & 4.71 & 79.80 & 1175.80                     \\
    & OrderEmb & 1.08 & \cellcolor{LightGreen} 2.12 & \cellcolor{LightGreen} 13.25 & \cellcolor{LightGreen} 122.81 & 1.21 & \cellcolor{LightGreen} 1.95 & \cellcolor{LightGreen} 7.90 & \cellcolor{LightGreen} 20.26 & 1.19 & \cellcolor{LightGreen} 2.94 & \cellcolor{LightGreen} 40.51 & \cellcolor{LightGreen} 194.11 \\
\bottomrule

\end{tabular}
\end{table*}

\begin{table*}[t]
\centering
\small
\caption{Q-error for MSCN on \dsb}\label{tab:qerr_MSCN_DSB}
\begin{tabular}{c|rrrr|rrrr|rrrr|rrrr}
\toprule
  \multirow{2}{*}{Algorithm} & \multicolumn{4}{c|}{Overall} & \multicolumn{4}{c|}{Template 018} & \multicolumn{4}{c|}{Template 027} & \multicolumn{4}{c}{Template 040} \\
  
   &  {50\%} & {75\%} & {95\%} & {99\%} & {50\%} & {75\%} & {95\%} & {99\%} & {50\%} & {75\%} & {95\%} & {99\%} & {50\%} & {75\%} & {95\%} & {99\%}  \\\midrule
     ERM  & 1.60 & 6.69  & 88.51 &  298.47 & 1.57 & 3.58 & 13.77 & 48.06 & 1.44 & 6.70 & 36.59 & 125.10 & 3.38 & 50.98 & 268.44 & 653.52\\ 
     \cline{1-17}  
     Deep CORAL  &  \cellcolor{LightGreen} 1.11 & \cellcolor{LightGreen} 3.51  & \cellcolor{LightGreen} 37.93  & \cellcolor{LightGreen}185.13 & \cellcolor{LightGreen} 1.05 & \cellcolor{LightGreen} 2.43 & \cellcolor{LightGreen} 8.09 & \cellcolor{LightGreen} 32.97 &  \cellcolor{LightGreen} 1.01 & \cellcolor{LightGreen} 3.82 & 33.39 & 123.72 & \cellcolor{LightGreen} 1.72 & \cellcolor{LightGreen} 10.26 & \cellcolor{LightGreen} 122.16 & \cellcolor{LightGreen} 385.06 \\
     DANN & 1.82 &  7.03 & 81.77 & 342.30  & 1.62 & 5.09 & 15.65 & \cellcolor{LightGreen}  27.11 & 4.45 & 17.03 & 237.67 & 778.32 & \cellcolor{LightGreen} 1.07 & \cellcolor{LightGreen} 11.18 &\cellcolor{LightGreen}  119.15 & \cellcolor{LightGreen} 278.67 \\
     Group DRO & 2.13 & 7.63 & \cellcolor{LightGreen} 51.75 & \cellcolor{LightGreen} 120.64 & 3.32 & 9.65 & 44.38 & 114.35 & \cellcolor{LightGreen} 1.11 & \cellcolor{LightGreen} 1.70 & \cellcolor{LightGreen} 6.77 & \cellcolor{LightGreen} 22.79 & 3.23 & \cellcolor{LightGreen} 15.02 & \cellcolor{LightGreen} 115.58 & \cellcolor{LightGreen} 127.94 \\
     Query Mixup & 2.03 & 5.67 & \cellcolor{LightGreen} 32.91 & \cellcolor{LightGreen} 111.27 & 1.95 & 5.07 & 24.88 & 54.26 & 1.45 & 5.63 & 65.50 & 247.54 & 3.32 & \cellcolor{LightGreen} 7.17 & \cellcolor{LightGreen} 31.44 & \cellcolor{LightGreen} 65.65 \\
     Query Mask & 1.48  & 5.90 & \cellcolor{LightGreen} 48.23 & \cellcolor{LightGreen} 162.41 & 1.26 & 3.01 & \cellcolor{LightGreen} 8.69 & \cellcolor{LightGreen} 20.46 & 1.77 & 7.70 & 51.08 & 172.86 & 3.34 & \cellcolor{LightGreen} 14.93 & \cellcolor{LightGreen} 89.37 & \cellcolor{LightGreen} 256.28\\
     OrderEmb  & \cellcolor{LightGreen} 1.05 & \cellcolor{LightGreen} 3.47 & 132.91 & 463.47  & \cellcolor{LightGreen} 1.05 & \cellcolor{LightGreen} 1.96 & \cellcolor{LightGreen} 7.14 & \cellcolor{LightGreen} 15.39 & 1.72 & 6.30 & 82.52 & 332.26 & \cellcolor{LightGreen} 1.64 & \cellcolor{LightGreen} 22.35 & 345.14 & 878.67 \\
\bottomrule

\end{tabular}
\end{table*}

\subsection{Training Efficiency}
\label{sec:exp:efficiency}

\begin{figure}[t]
 \centering
 \begin{tabular}[t]{c}
 \includegraphics[width=0.7\columnwidth]{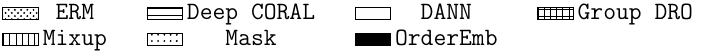}
 \vspace*{-0.2cm} 
\end{tabular}
	\begin{tabular}[h]{c}
        \subfigure[MLP] {
				\includegraphics[width=0.48\columnwidth]{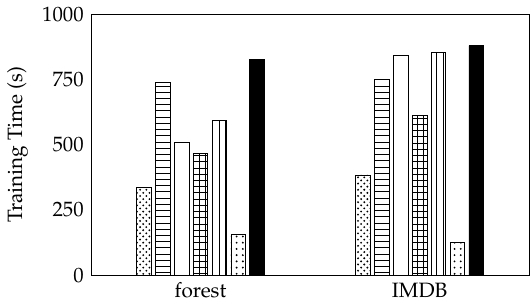}
			\label{fig:train_time:dnn}
		} 
        \subfigure[MSCN] {
			\includegraphics[width=0.48\columnwidth]{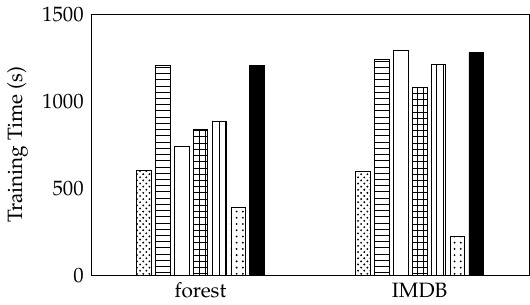}
			\label{fig:train_time:mscn}
		}
	\end{tabular}
	\caption{ Training Time (Seconds)}
        \label{fig:training_time}
\end{figure}

We compare the training time of the 6 algorithms and the ERM in \CardOOD, where the results are shown in Fig.~\ref{fig:train_time:dnn} for MLP and Fig.~\ref{fig:train_time:mscn} for MSCN respectively. 
It is worth mentioning that the prediction time of the models trained by different algorithms is the same, given the same neural network architecture.
Compared to ERM, all the algorithms are slower excpet Query Masking which randomly drops query predicates. 
In general, Deep CORAL and OrderEmb take the longest time for training. The bottleneck of Deep CORAL is the computation and alignment of two covariance matrices in each gradient step. 
For OrderEmb, online sampling of the contrastive query set incurs high computational overhead. The training efficiency of DANN, Group DRO and Mixup is in the middle in \CardOOD.  

\subsection{Sensitivity Analysis}
\label{sec:exp:sensitivity} 

\subsubsection{Data Efficiency}
\begin{figure}[t]
    \centering
    \includegraphics[width=0.3\columnwidth]{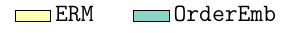} \\
	\begin{tabular}[h]{c}
		\subfigure[MLP on \forest] {
        \includegraphics[width=0.48\columnwidth]{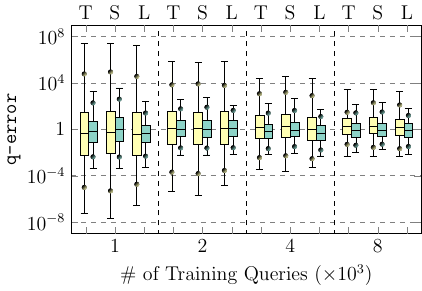}
			\label{fig:data:forest_dnn}
		}
		\subfigure[MSCN on \forest] {
			\includegraphics[width=0.48\columnwidth]{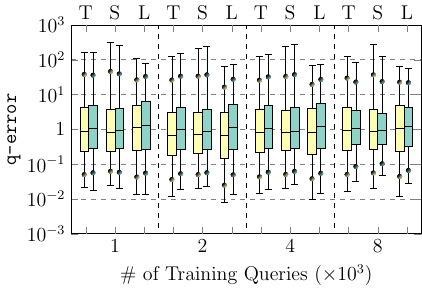}
			\label{fig:rowbookpre}
            }\\
             \subfigure[MLP on \imdb] {
        \includegraphics[width=0.48\columnwidth]{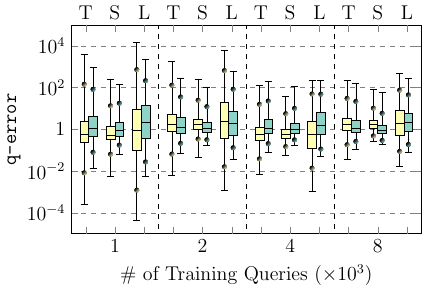}
			\label{fig:layerndcg}
		} 
        \subfigure[MSCN on \imdb] {
			\includegraphics[width=0.48\columnwidth]{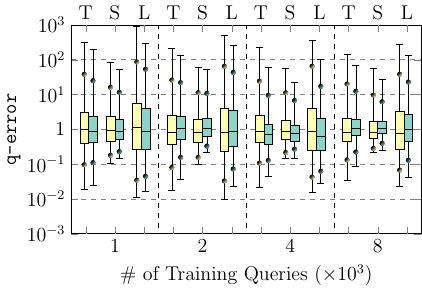}
			\label{fig:rowbookndcg}
		}
	\end{tabular}
	\caption{Varying \# of Training Queries}
        \label{fig:data_efficiency}
\end{figure}

We investigate for cardinality estimation, whether \OrderEmb is more data efficient than ERM, given limited training queries. Here, we keep a balanced training query workload with uniformly distributed selection and join conditions, but vary the size of the training set in \{1000, 2000, 4000, 8000\}. 
Fig.~\ref{fig:data_efficiency} presents the prediction \Qerror of ERM and \OrderEmb on \forest and \imdb, where MLP and MSCN serve as the learned estimators. In Fig.~\ref{fig:data_efficiency}, the annotations `T', `S', `L' denotes the Total, Small and Large test queries, and `$\bullet$' on the standard box plot denotes the $95\%$ quantiles of the estimations. The accuracy of OrderEmb consistently surpasses ERM regardless of the model types and the number of training queries. 
As the number of training queries decreases from 8,000 to 1,000, OrderEmb demonstrates performance robustness, indicating its data efficiency. These results validate the utilization of the partial order of the sampled contrastive queries for cardinality estimation.

\begin{figure}[h]
 \centering
 \begin{tabular}[t]{c}
 \includegraphics[width=0.4\columnwidth]{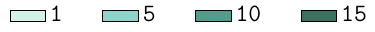}
 \vspace*{-0.2cm} 
\end{tabular}
	\begin{tabular}[h]{c}
        \subfigure[MLP on \imdb] {
				\includegraphics[width=0.48\columnwidth]{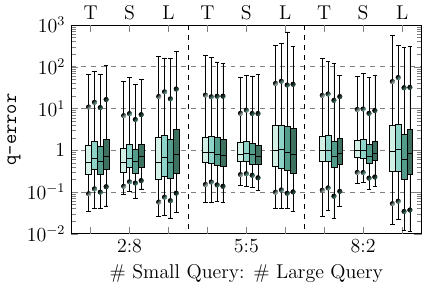}
			\label{fig:varying_negs:dnn}
		} 
        \subfigure[MSCN on \imdb] {
			\includegraphics[width=0.48\columnwidth]{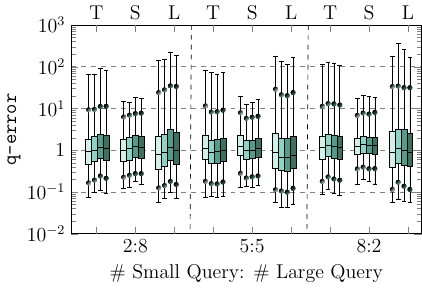}
			\label{fig:varying_negs:mscn}
		}
	\end{tabular}
	\caption{Varying \# of Contrastive Queries in OrderEmbed}
        \label{fig:varying_negs}
\end{figure}

\begin{figure*}[htbp]
 \centering 
\includegraphics[width=0.9\textwidth]
            {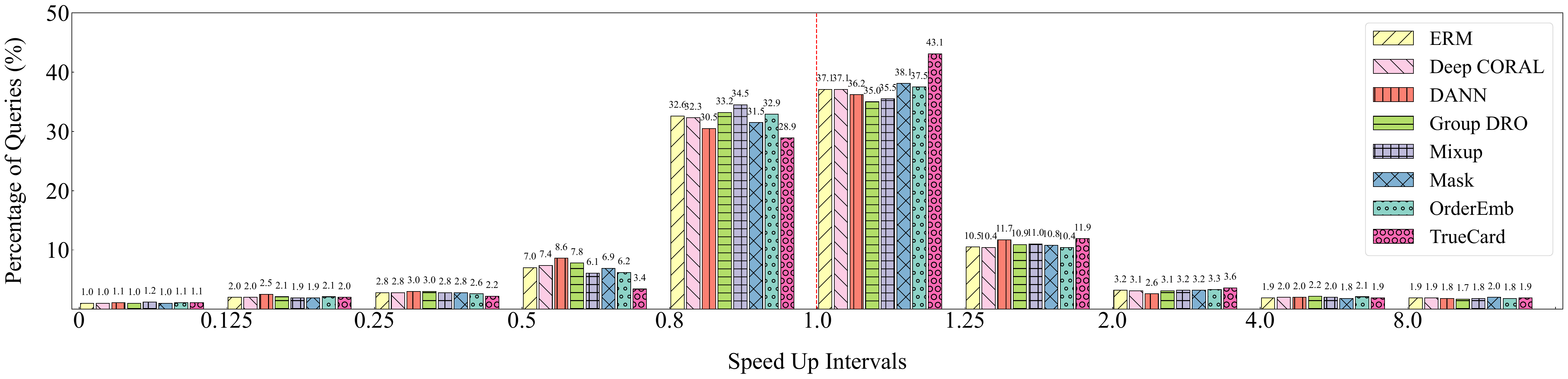}
 \caption{Query Speedup in \PostgreSQL for \kw{IMDB}: \# Table in the Test Queries is OOD}
 \label{fig:table_num_speedup_imdb}
\end{figure*}

%
%

\begin{figure*}[htbp]
 \centering
 \begin{tabular}[h]{c}
 \subfigure[Query Speedup for \dsb] {
\includegraphics[width=0.54\textwidth]
                {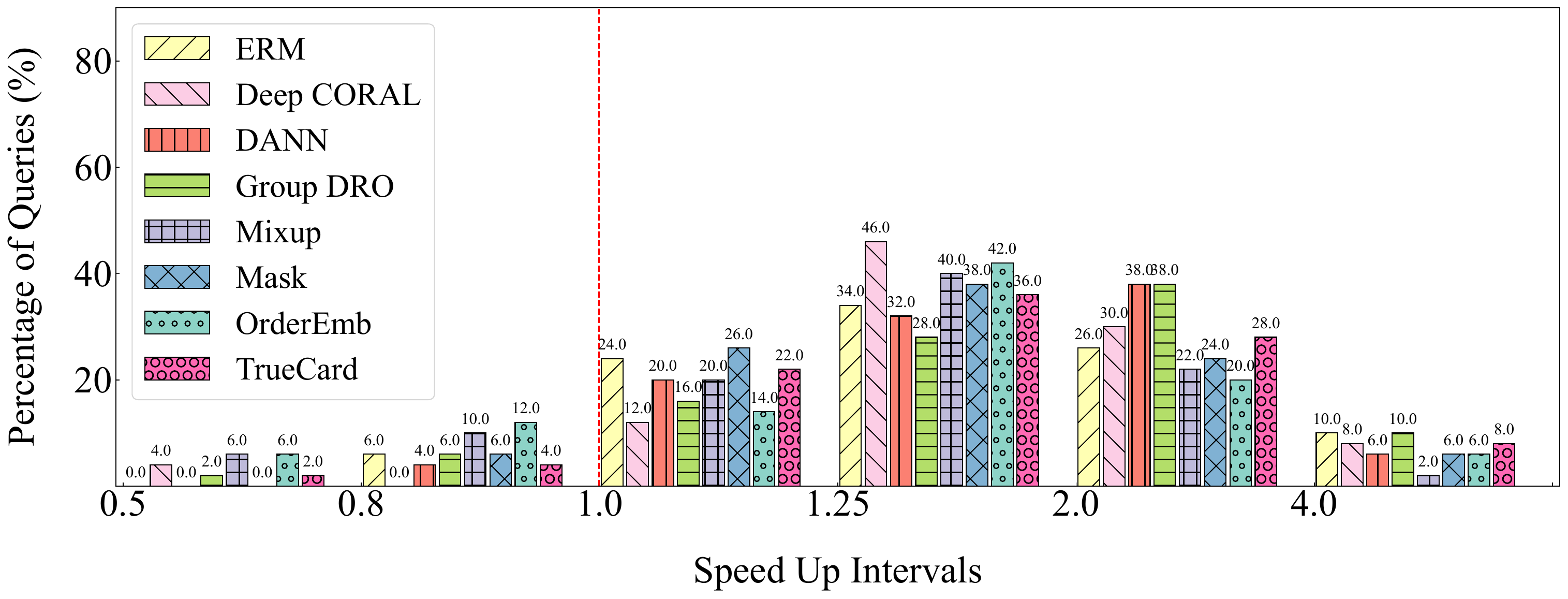}
 \label{fig:table_num_speedup_dsb}
 }
 \subfigure[Query Speedup for \job ] {\includegraphics[width=0.36\textwidth]
                {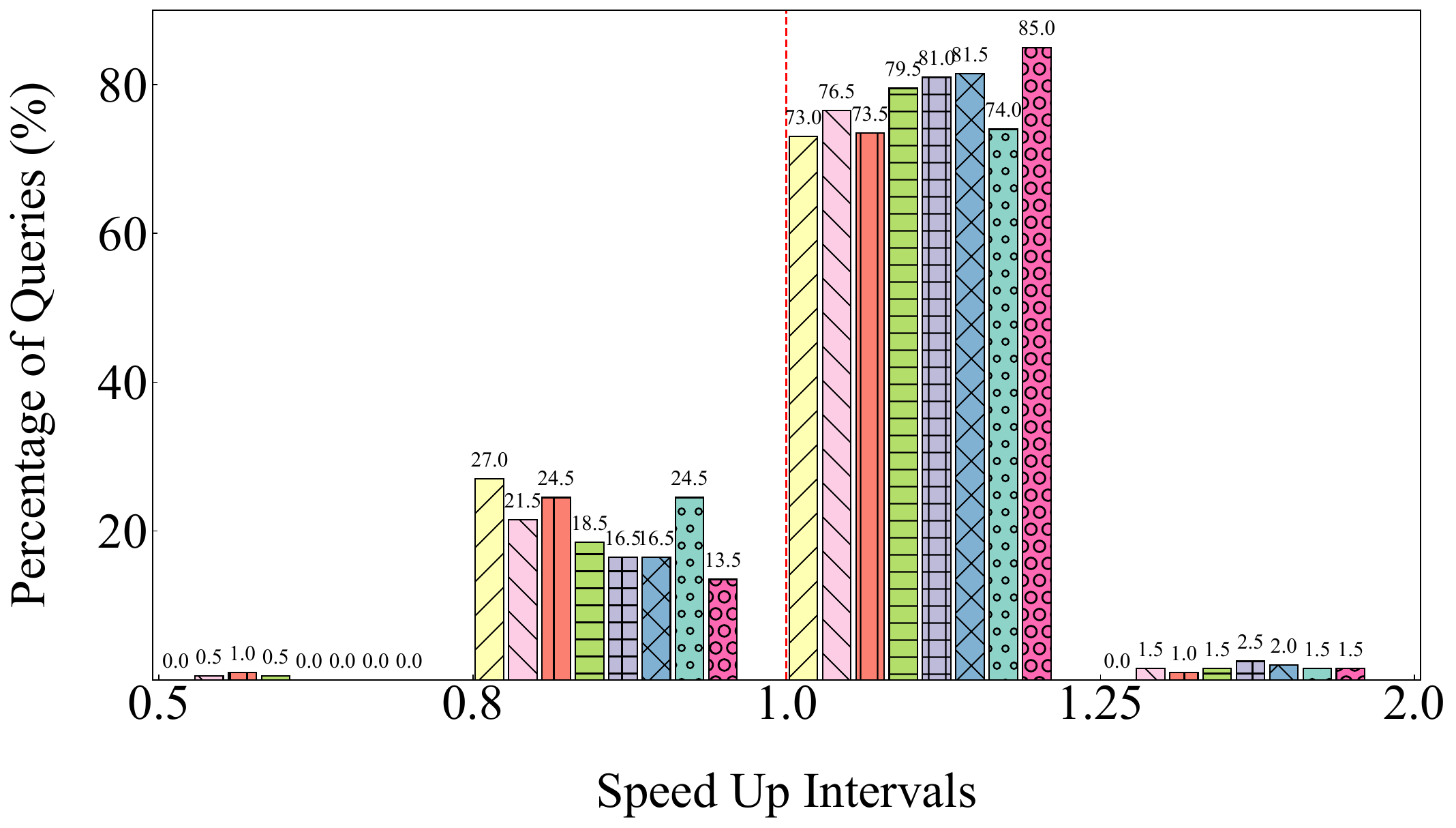}
                \label{fig:table_num_speedup_joblight}
 }
 \end{tabular}
 \vspace{-0.1cm}
\caption{Query Speedup in \PostgreSQL for \dsb and \job: Test Query Template is OOD}
 \vspace{-0.2cm}
\end{figure*}

\begin{figure}[htbp]
 \centering

\end{figure}

\comment{
\begin{figure*}[htbp]
  \centering
	\begin{tabular}[h]{c}
        \subfigure[Speedup in \PostgreSQL for \kw{IMDB}: \# Table in the Test Queries is OOD]{
            \includegraphics[width=0.9\textwidth]
            {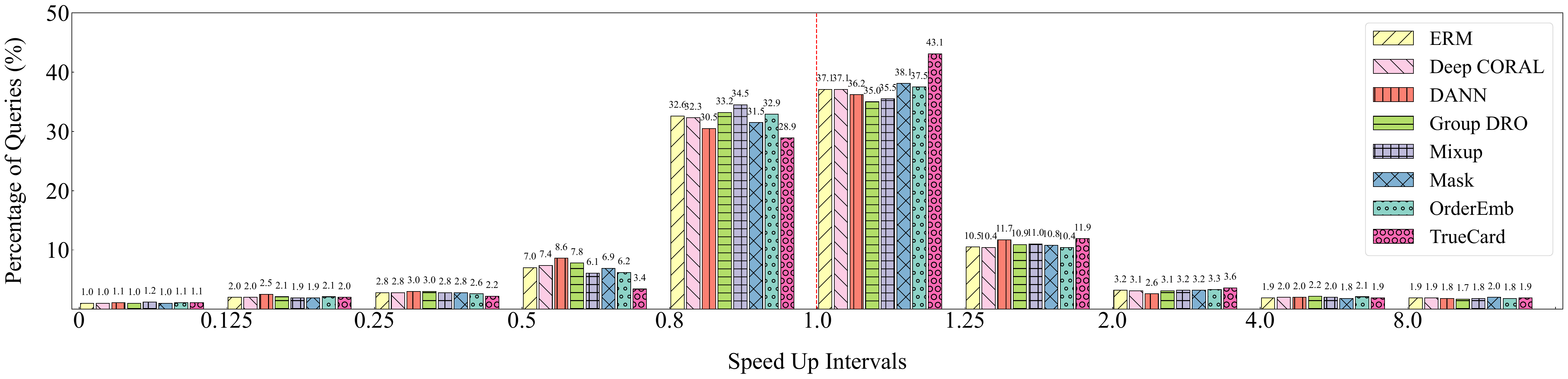}
                \label{fig:table_num_speedup_imdb} } \\

          \subfigure[Speedup in \PostgreSQL for \job: Test Query Template is OOD]{
                \includegraphics[width=0.45\textwidth]
                {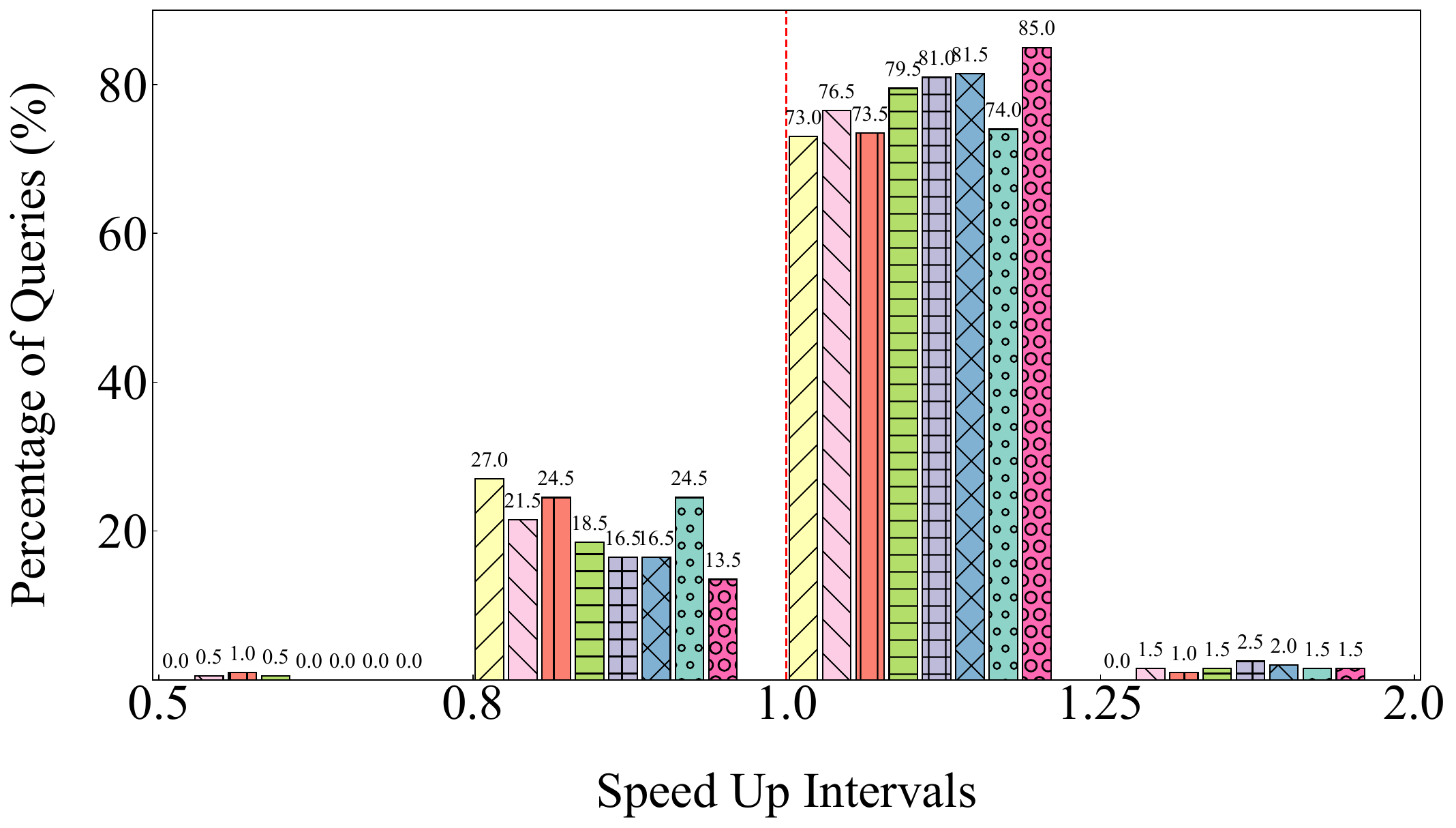}
                \label{fig:table_num_speedup_joblight} } \\ 
                
            \subfigure[Speedup in \PostgreSQL for \dsb: Test Query Template is OOD]{
                \includegraphics[width=0.9\textwidth]
                {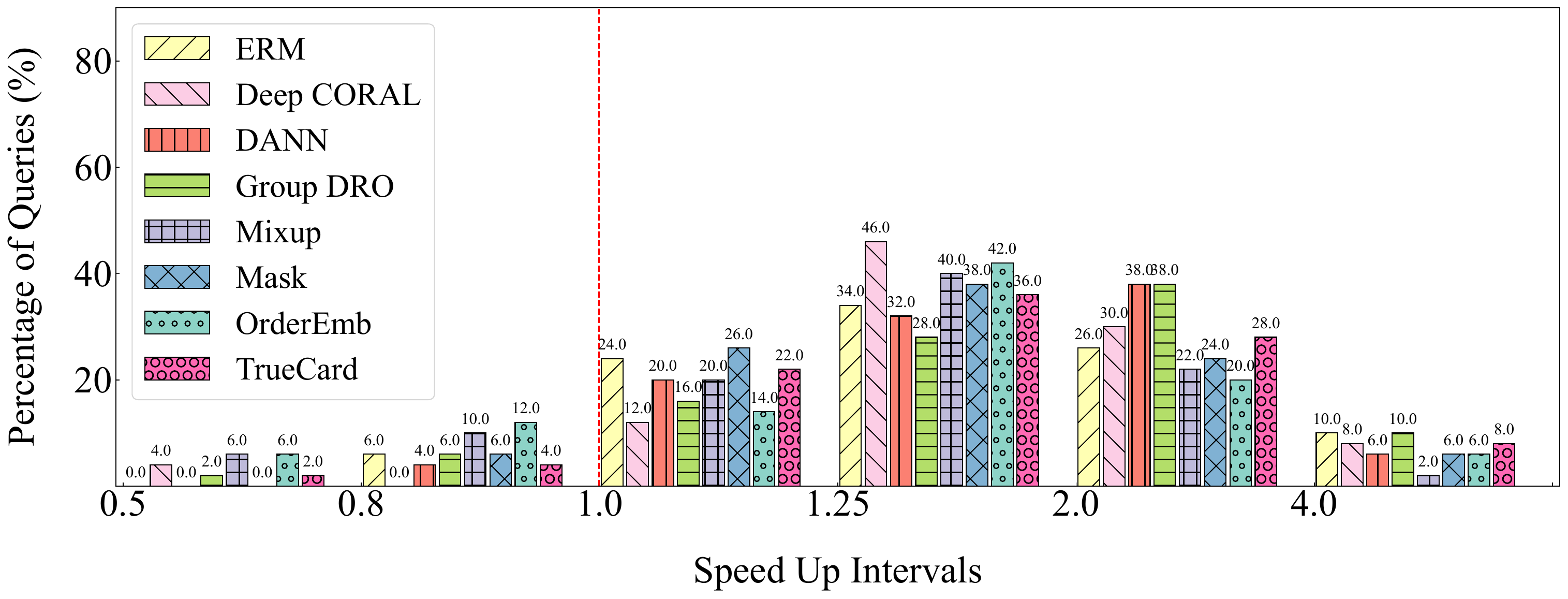}
                \label{fig:table_num_speedup_dsb} } 
	\end{tabular}
  \caption{Query Speedup for \CardOOD in \PostgreSQL }
  \label{fig:main}
\end{figure*}
}

\subsubsection{Impact of the number of contrastive queries}
For the algorithm \OrderEmb, we study how the number of contrastive queries, i.e., $|\mathcal{C}(q)|$ in Algorithm~\ref{alg:order},  influences the prediction \Qerror. 
Fig.~\ref{fig:varying_negs} illustrates the \Qerror on \imdb under imbalance training workloads when varying $|\mathcal{C}(q)|$ in \{1, 5, 10, 15\}. 
We find that when the number is set to too large (e.g., 10, 15) or too small (e.g., 1), the prediction accuracy degenerates slightly. Since if $|\mathcal{C}(q)|$ is too small, the model cannot fully utilizes the order information of the sampled queries. On the contrary, the training objective will focus on fitting complex order constraints, neglecting the \Qerror to be minimized. It is worth to noting that the order embedding is an auxiliary learning task to alleviate OOD. A strict query order embedding is a complicated learning objective and may not necessarily contribute to accurate predictions. 

\subsection{Query Optimization in PostgreSQL}
We show the performance of \CardOOD in \PostgreSQL by
injecting the cardinality predictions of different models into the query optimizer.
 We perform experiments to evaluate two OOD scenarios on join queries as follows:

\stitle{\# Tables in Test Queries is OOD.} We evaluate the  execution of queries with different numbers of tables from that of the training queries.  We collect a set of queries with the number of tables ranging from $1$ to $5$ on \imdb.
For each number, we generate $9,000$ queries involving both PK/FK and FK/FK joins.  
We train the models using queries with $1$ to $3$ tables and evaluate the end-to-end execution time using queries with $5$
tables. 
As optimizing $5$-table queries requires the estimated cardinalities of $4-$table \textit{sub-plans}, which are unseen in the 
training set.

Table~\ref{tbl:query_time} lists the total query execution time for all the models on \imdb, as well as the ratio of the total time compared to \PostgreSQL built-in estimator.
The algorithms with the best and second-best performance are marked as bold and highlighted, respectively. 
Overall, the running time can be reduced by up to $5.6\%$ (Deep CORAL) 
by using the estimations from learned models. Among these models, DANN performs
worse than others, which is consistent with the performance of \Qerror shown
in Table \ref{tab:qerr_MSCN_IMDB}, where DANN achieves a median of $2.09$ for complex queries
under simple/complex ratio of  0.2. It is worth noting that the \Qerror is not necessarily 
consistent with the query execution time. For example, Group DRO has a 
worse  \Qerror in most cases compared with other models but achieves nearly the 
shortest execution time. 
Fig.~\ref{fig:table_num_speedup_imdb} illustrates the speedup over \PostgreSQL for all models as well as the true cardinalies (True Card). Here, we partition the queries into several  
speedup intervals and plot the percentages of queries felling in each interval. For the queries with a speedup larger than $1$, the learned estimators perform better than \PostgreSQL built-in estimator and vice versa. 
In general, the learned estimators demonstrate superior performance compared to
\PostgreSQL, outperforming it on approximately $53\%$ of queries for Group DRO and $56\%$ for Query Masking. About 90\% of queries fell into the speedup 
interval of $0.5 \sim 2.0$. 
We also observe some cases where the True Card leads to performance regression in the speedup interval $0.125$ to $1.0$, revealing the 
intricacy of query optimization and execution. 

\stitle{Test Query Template is OOD.} 
On \job and \dsb, we evaluate the queries with different templates from those used to train models. 
Specifically, For \dsb, we follow the same setting to \cref{sec:exp:robust}. For \job, we use $60$ templates
to train the models and test the queries from template $t70$ on \PostgreSQL, as it contains $5$ tables and has a larger 
search space on sub-plans.

Table~\ref{tbl:query_time} shows the overall running time, and
Fig.~\ref{fig:table_num_speedup_dsb}  depicts the speedup in \PostgreSQL for all the models on \dsb. 
Overall, the learning strategies achieve $26\%$ to $36\%$ 
improvement over \PostgreSQL. Almost all the learning strategies in \CardOOD 
outperform ERM, except Query Mixup.
We notice that the plans from True Card 
achieve nearly $36\%$ improvement over \PostgreSQL, which is more remarkable than the improvement in \imdb and \job. 
This significant improvement is consistent with the findings 
in the original paper \cite{DBLP:conf/dsb/ding2021}, where 
injecting True Card into {\sl SQL Server} can reduce elapsed time by up to
$41\%$, compared to the built-in estimator. 
We also observe that the plans produced by some learning strategies are
slightly faster than those produced by True Card.
The results indicate that the plans 
generated by True Card are optimal w.r.t. the cost model, 
but not necessarily optimal in a  real execution environment. 

Fig.~\ref{fig:table_num_speedup_joblight} illustrates the speedup in  \PostgreSQL for all
the models on \job and the summary of running time is in Table~\ref{tbl:query_time}.
Compared with \PostgreSQL, the running time is reduced by up to $4.5\%$ for
Query Mixup. We observe that ERM only outperforms \PostgreSQL slightly, which
is  different from the scenario where the number of tables in the test query is OOD. The reason
may be that the model is trained on only $60$ templates on \job, which is far less than \imdb. Such a scenario underlines the significance of specialized learning strategies for OOD cases. 
Group DRO and Mixup perform similarly to the best competitor, Query Masking, 
indicating the effectiveness of the learning strategy. OrderEmb does not achieve the best performance but is still better than the plain training strategy, ERM. 
Fig.~\ref{fig:table_num_speedup_joblight} shows that the model outperforms \PostgreSQL on over 70\% of queries in total. 

\comment{
\ruimodify{For DSB, we use $10$ templates to train the model and $3$
templates to evaluate. The evaluation templates are randomly sampled 
from those with $5$ tables to align with the experimental settings for 
\imdb and \job. 
For each template, we generate $100$ queries.

Table~\ref{tbl:query_time} shows the overall running time, and
Fig.~\ref{fig:table_num_speedup_dsb}  shows the speedup in \PostgreSQL for all the models. 
Overall, the learning strategies achieves $26\%$ to $36\%$ 
improvement over $\PostgreSQL$. Almost all the robust learning strategies 
outperform ERM, except Query Mixup.
We notice that plans with true cardinality 
achieve nearly $36\%$ improvement over \PostgreSQL, which is much better
compared to \imdb and \job. 
This significant improvement can be explained by the experiment finding 
in the original \dsb paper \cite{DBLP:conf/dsb/ding2021}, where 
injecting true cardinalities into SQL Server can reduce elapsed time by up to
$41\%$,compared to the native cardinality estimator. 
We also observe that plans produced by some learning strategies can be 
slightly better than the plan with true cardinality.
The reason is that the plan 
generated by true cardinality is optimal with respect to the cost model, 
not necessarily optimal in the real experimental environment. 
}
}

\comment{
\begin{figure*}[htbp]
  \centering
	\begin{tabular}[h]{c}
        \subfigure[ttt] {
				\includegraphics[width=0.28\columnwidth]{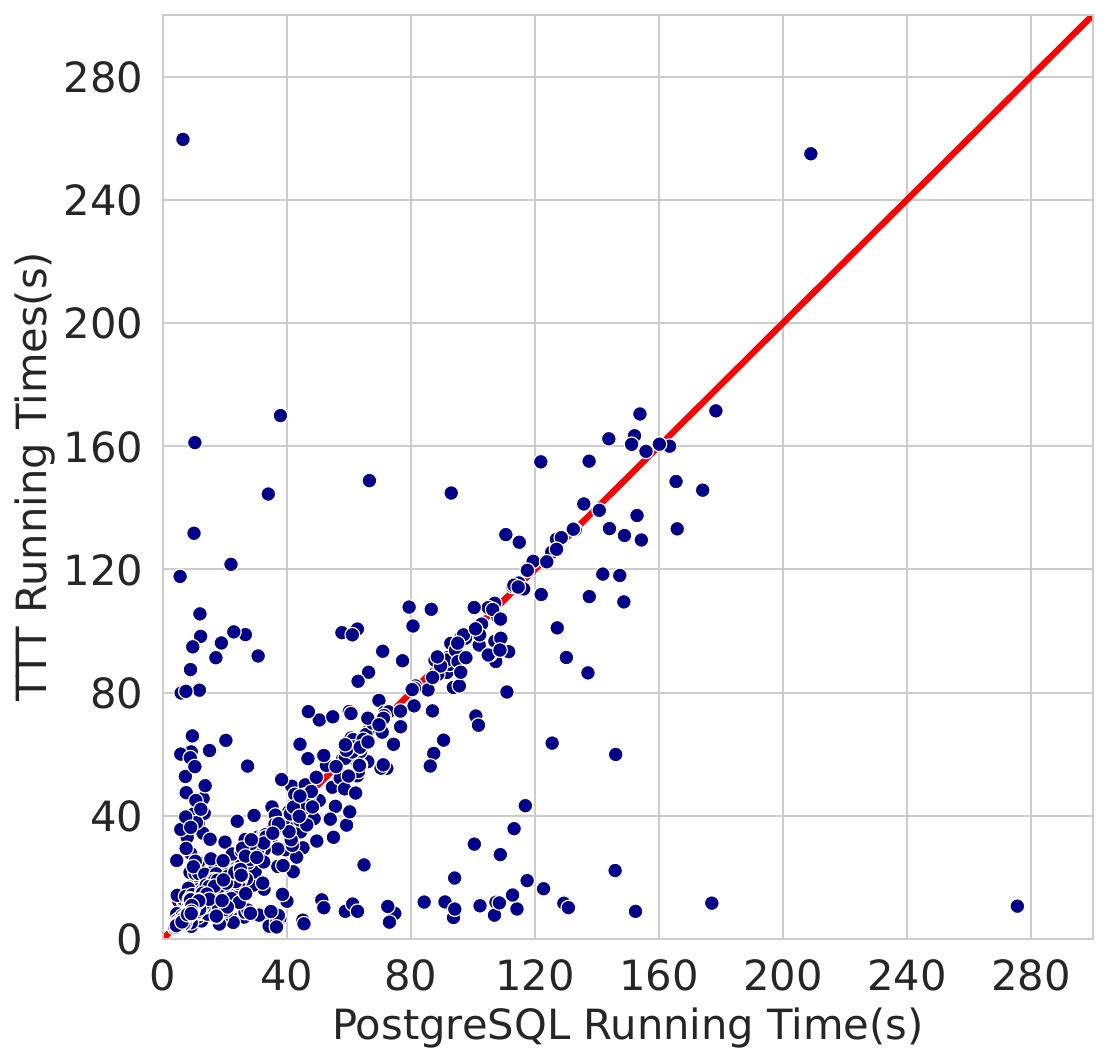}
			\label{fig:time_ttt_pg}
		} 
          \subfigure[erm] {
				\includegraphics[width=0.28\columnwidth]{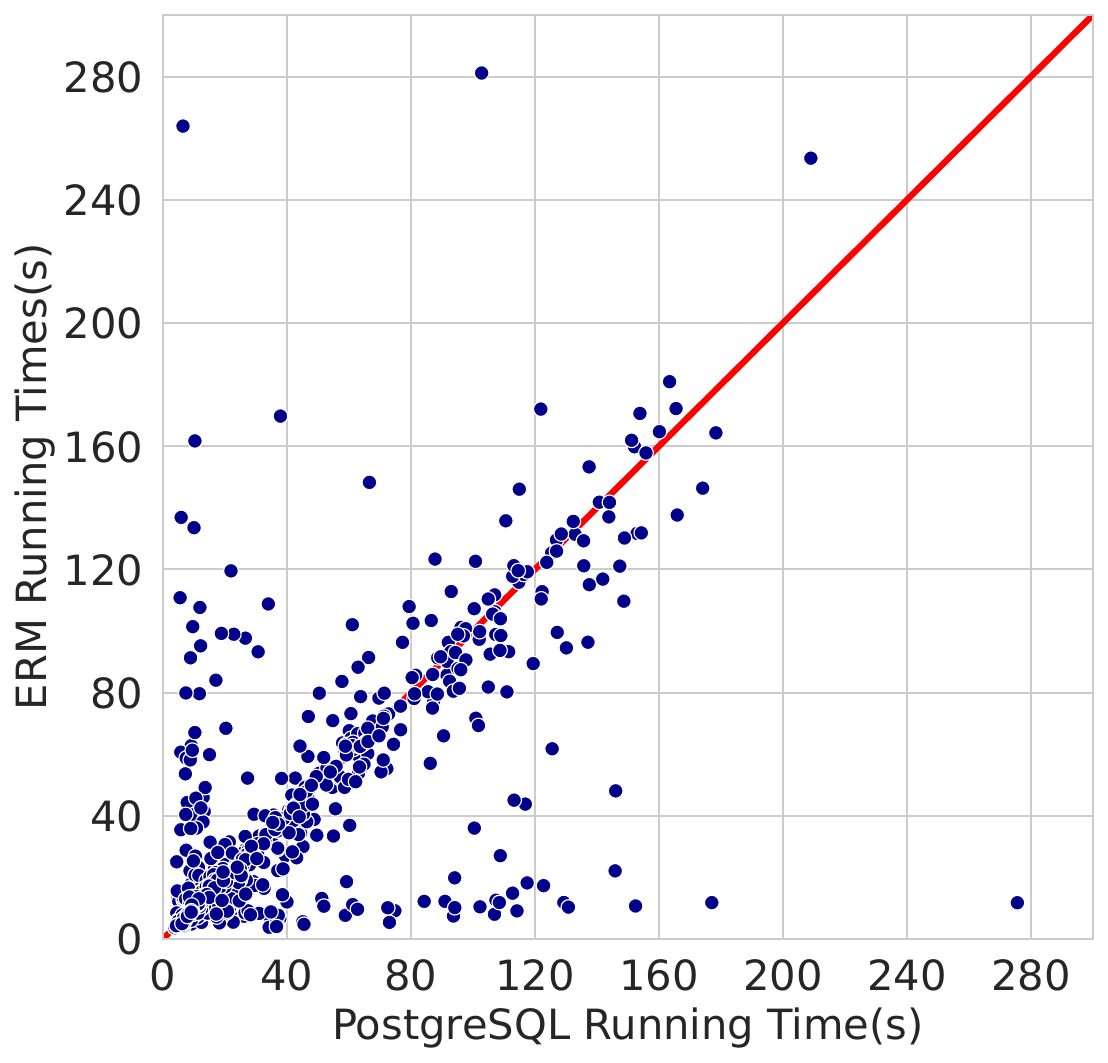}
			\label{fig:time_erm_pg}
		} 
          \subfigure[coral] {
				\includegraphics[width=0.28\columnwidth]{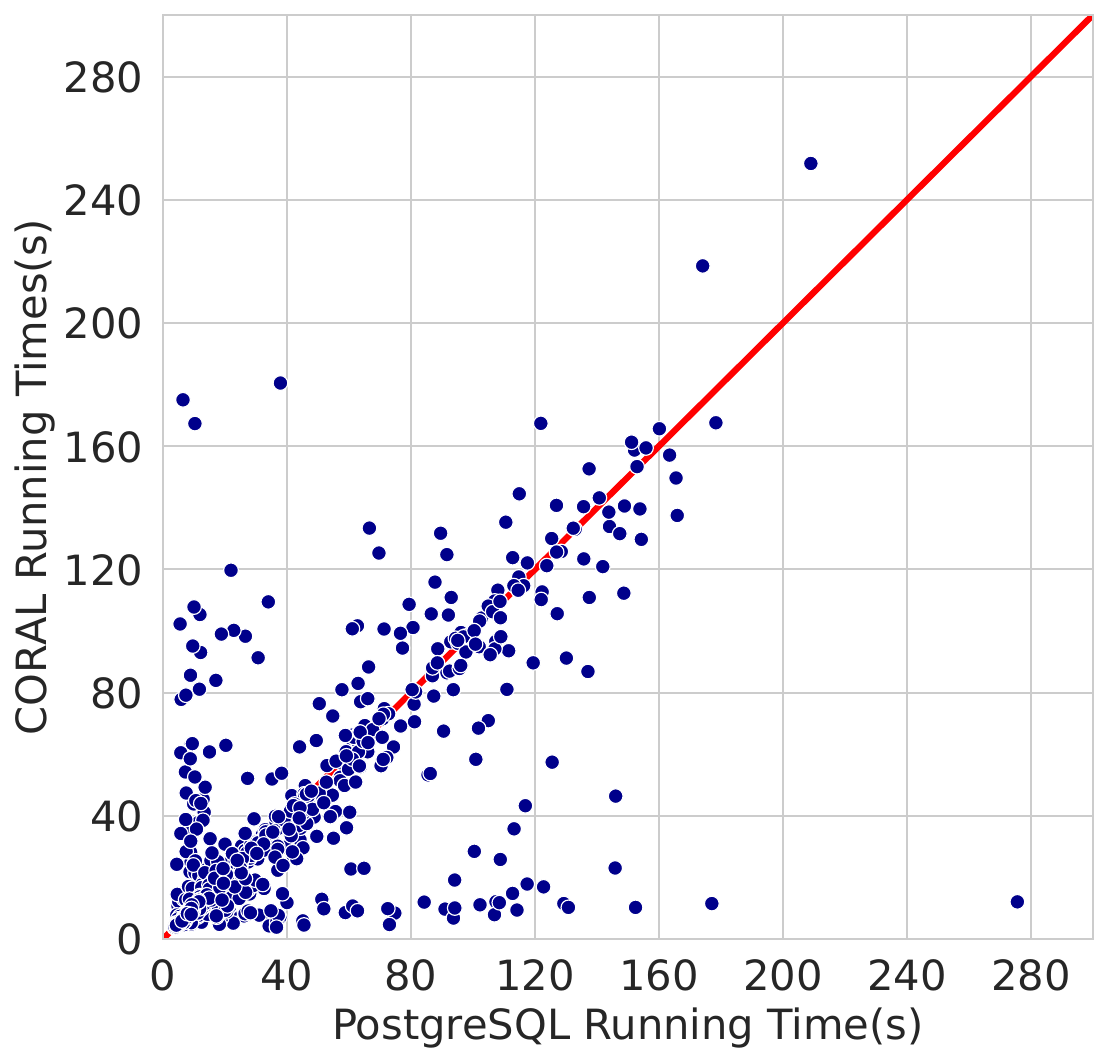}
			\label{fig:time_coral_pg}
		} 
          \subfigure[dann] {
				\includegraphics[width=0.28\columnwidth]{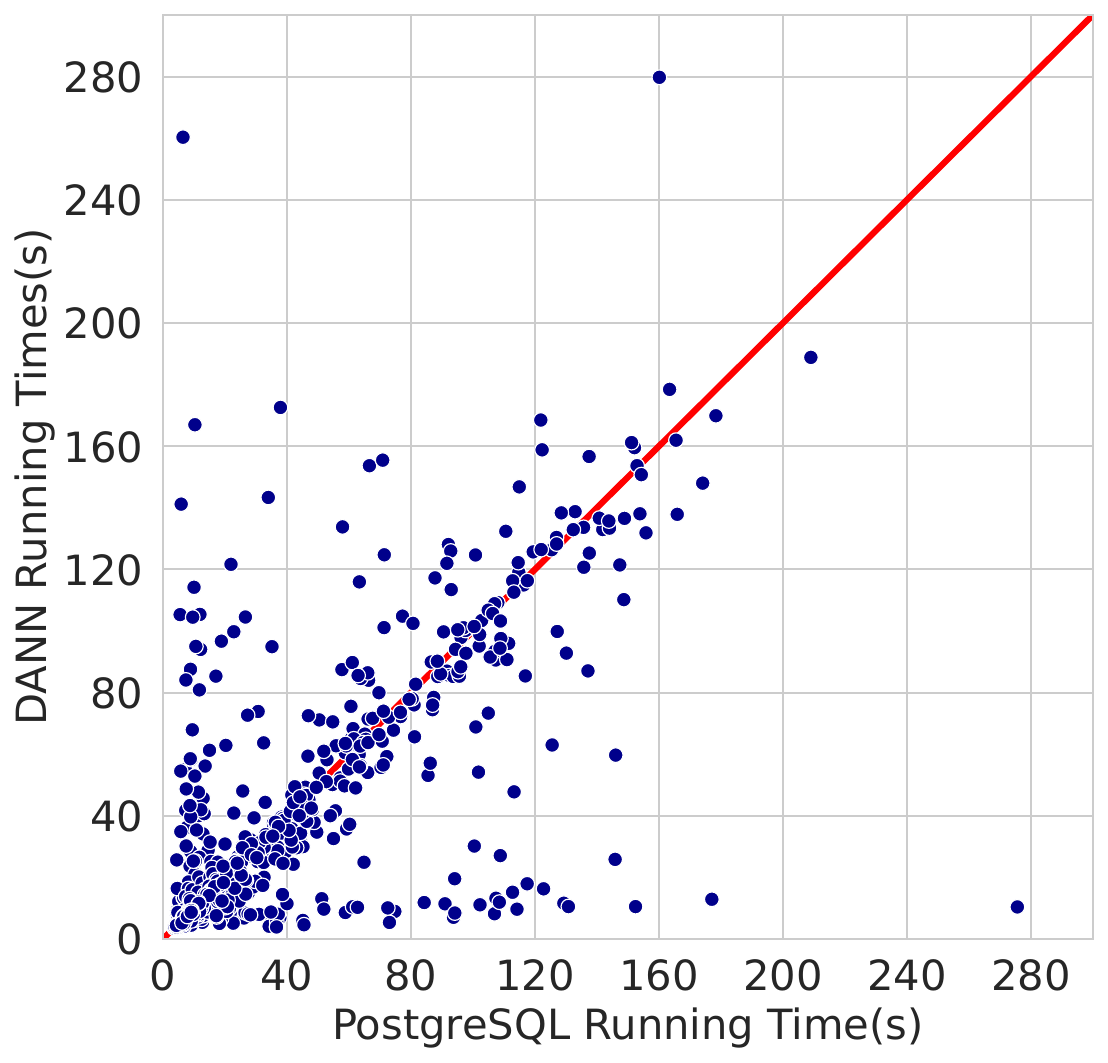}
			\label{fig:time_dann_pg}
		} 
          \subfigure[groupdro] {
				\includegraphics[width=0.28\columnwidth]{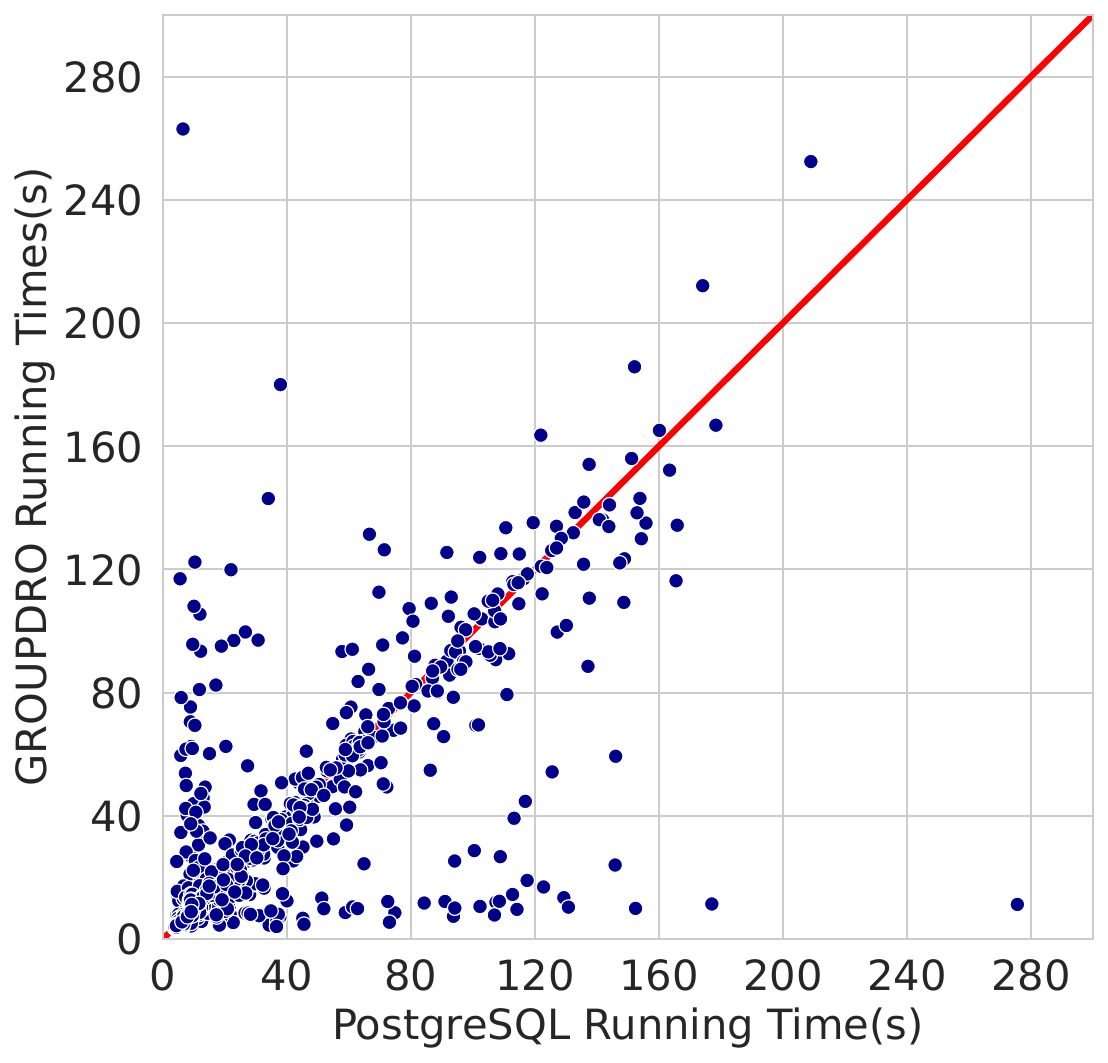}
			\label{fig:time_groupdro_pg}
		} 
          \subfigure[mixup] {
				\includegraphics[width=0.28\columnwidth]{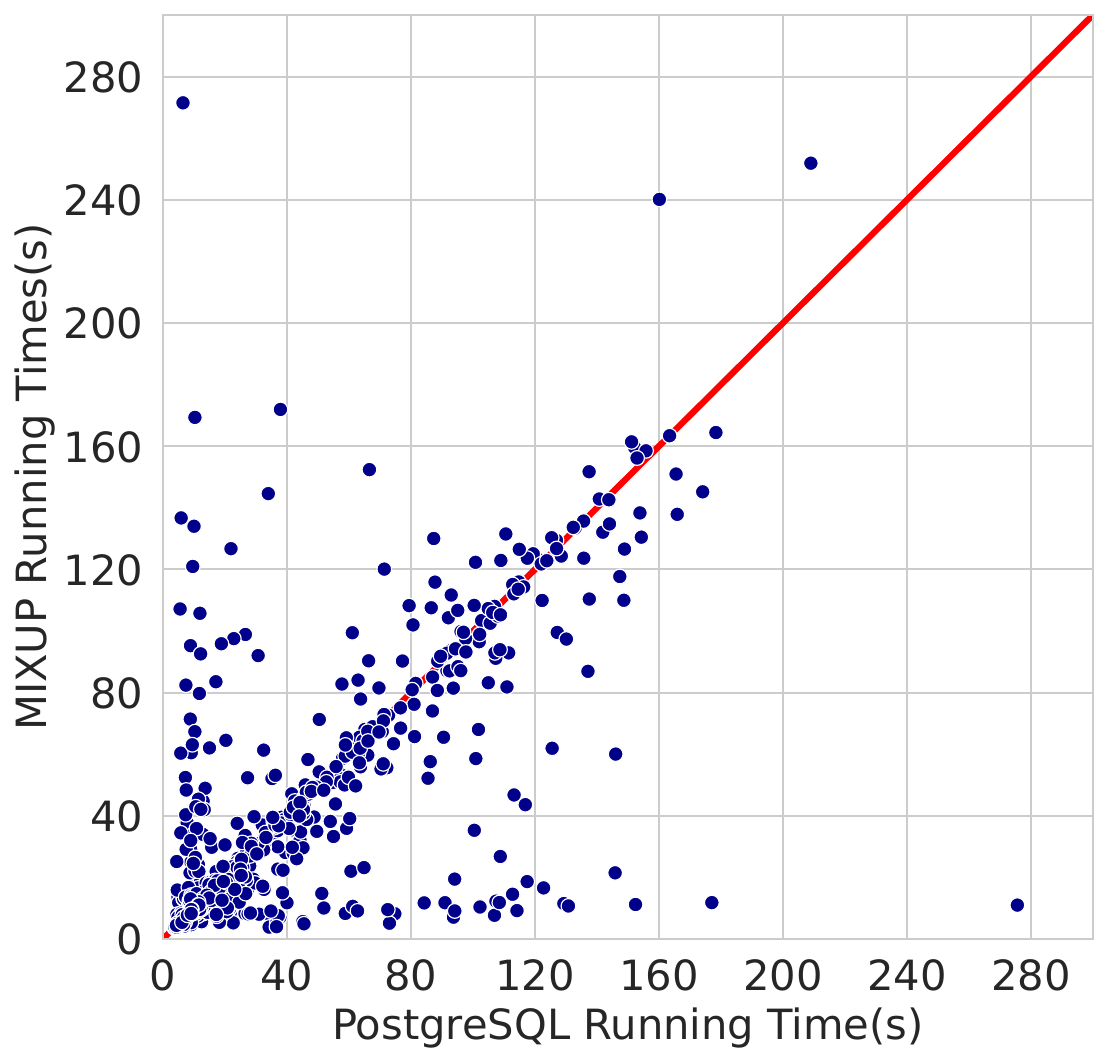}
			\label{fig:time_mixup_pg}
		} 
          \subfigure[mask] {
				\includegraphics[width=0.28\columnwidth]{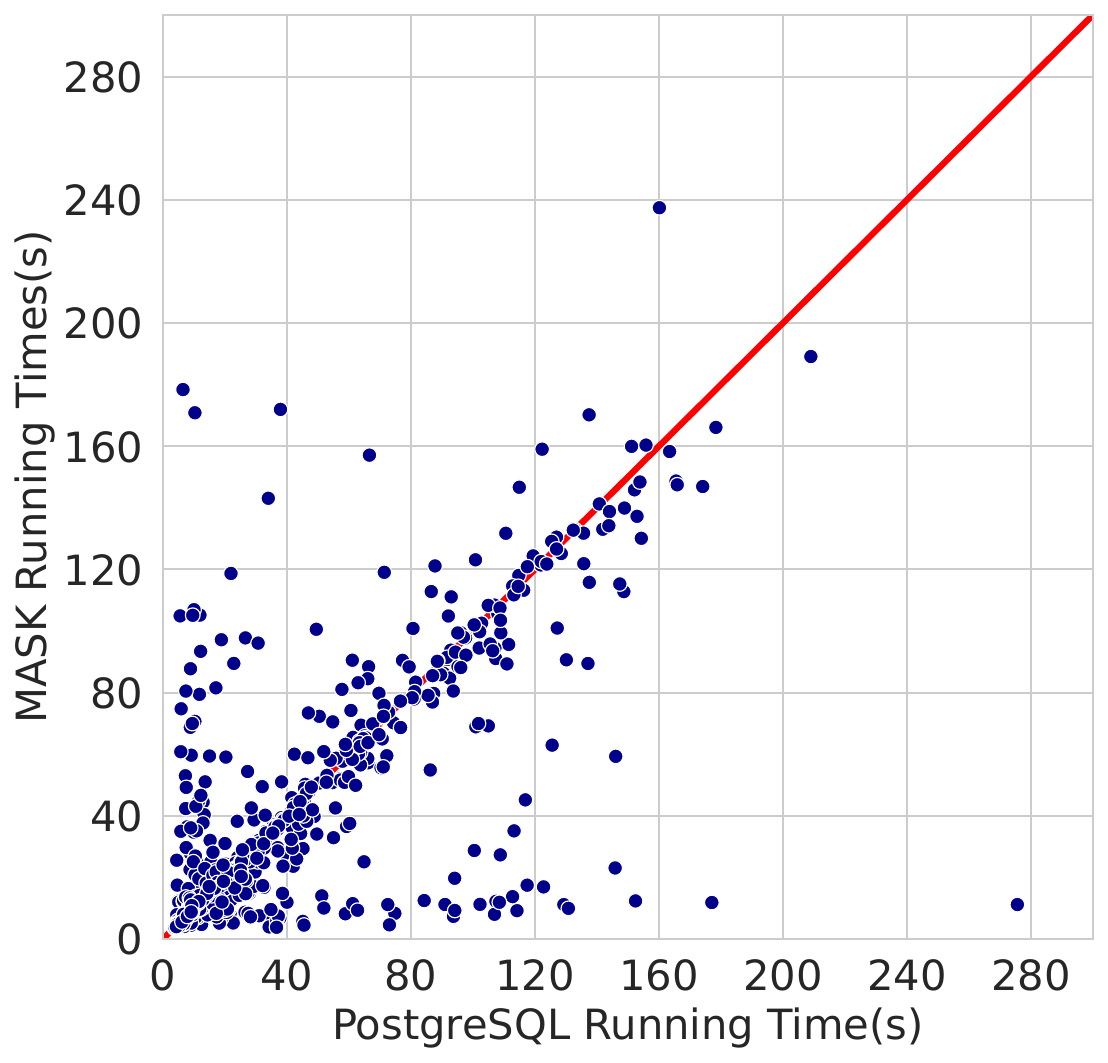}
			\label{fig:time_mask_pg}
		} 
	\end{tabular}

  \caption{Time comparison}
  \label{fig:main}
\end{figure*}
}

\comment{
\begin{figure*}[htbp]
  \centering
  
	\begin{tabular}[h]{c}
        \subfigure[ttt] {
				\includegraphics[width=0.28\columnwidth]{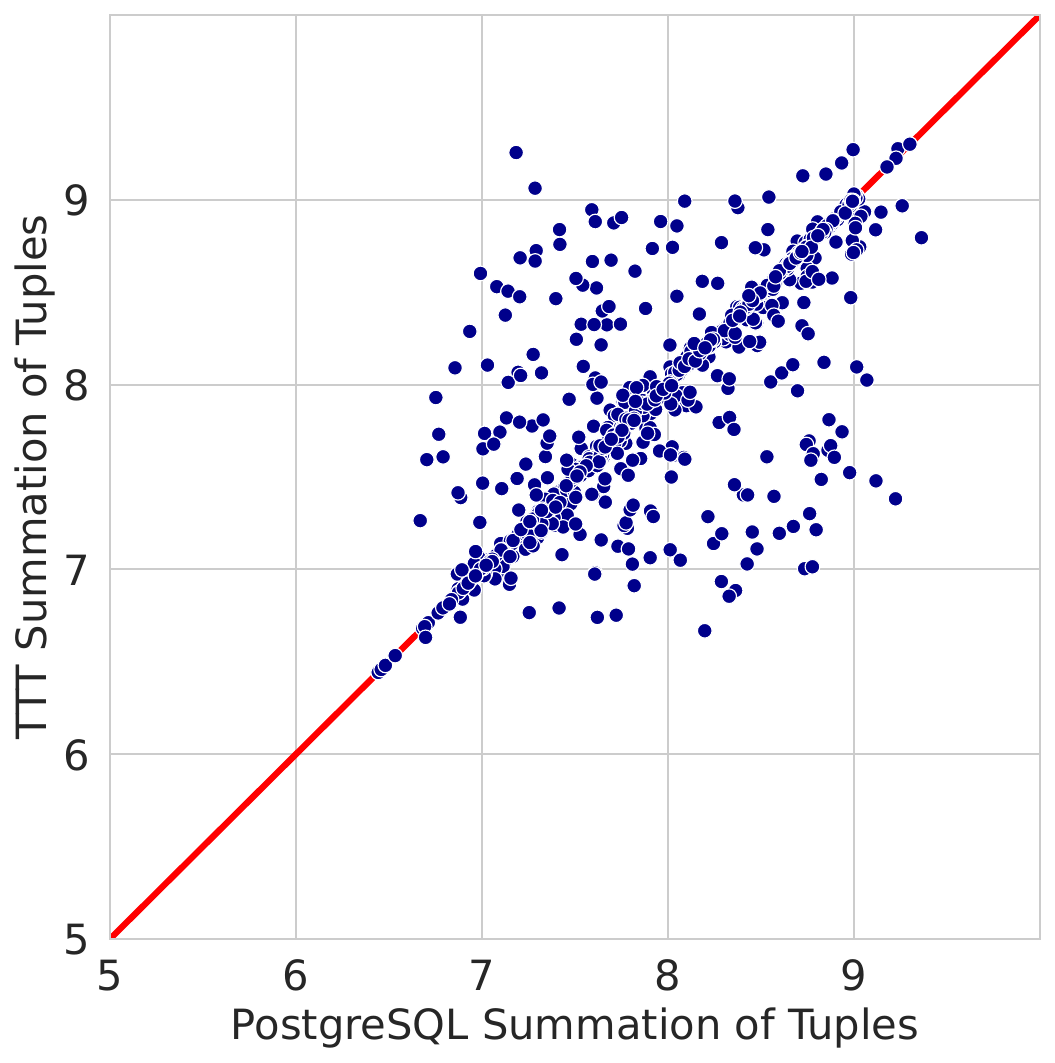}
			\label{fig:tuple_ttt_pg}
		} 
          \subfigure[erm] {
				\includegraphics[width=0.28\columnwidth]{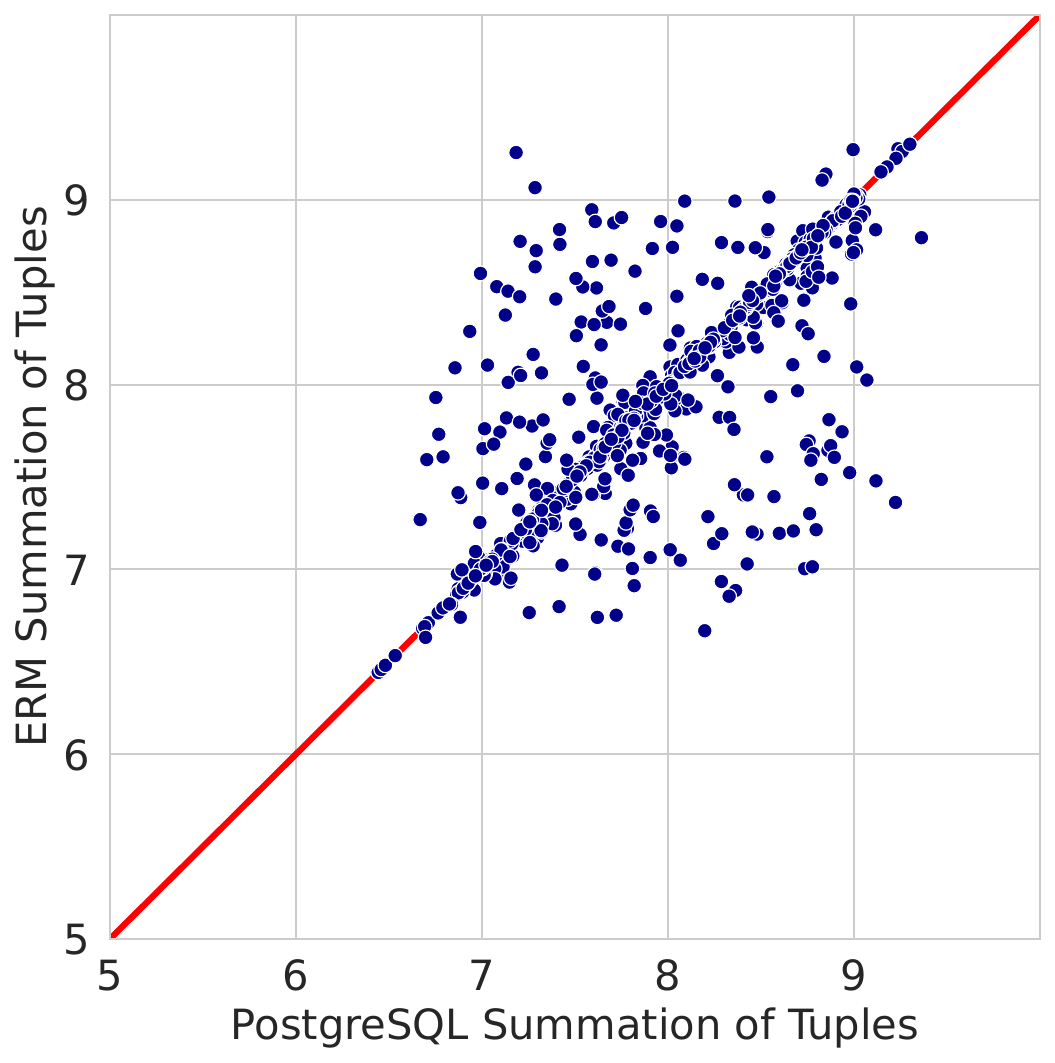}
			\label{fig:tuple_erm_pg}
		} 
          \subfigure[coral] {
				\includegraphics[width=0.28\columnwidth]{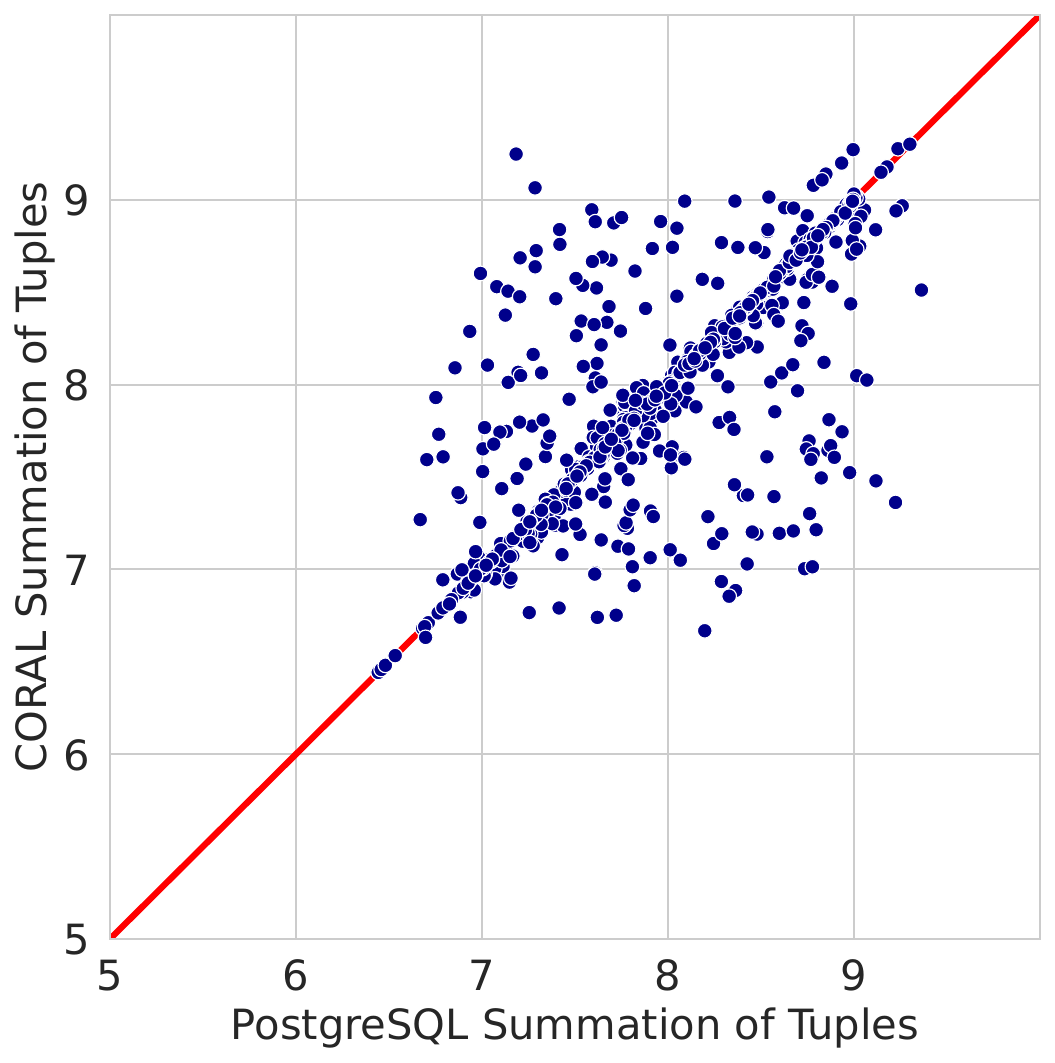}
			\label{fig:tuple_coral_pg}
		} 
          \subfigure[dann] {
				\includegraphics[width=0.28\columnwidth]{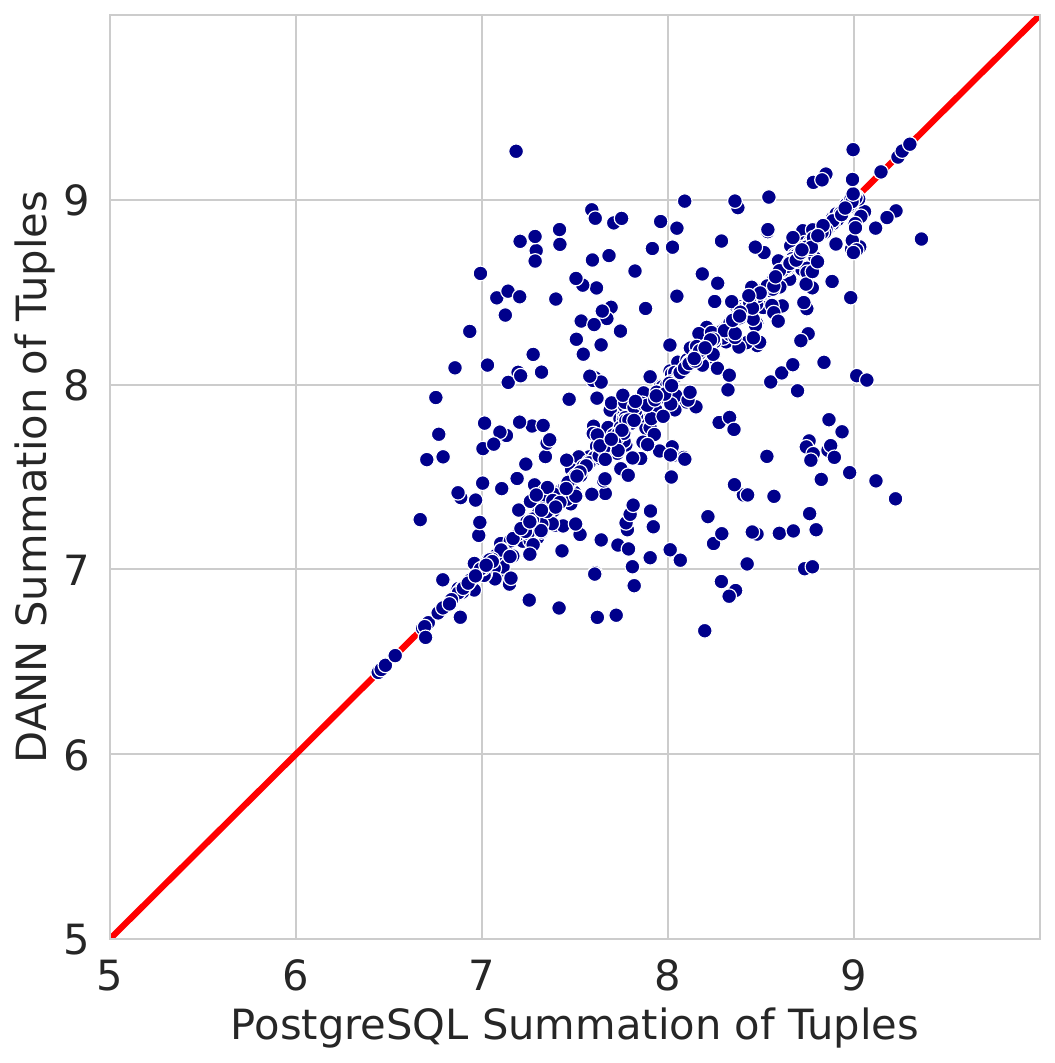}
			\label{fig:tuple_dann_pg}
		} 
          \subfigure[groupdro] {
				\includegraphics[width=0.28\columnwidth]{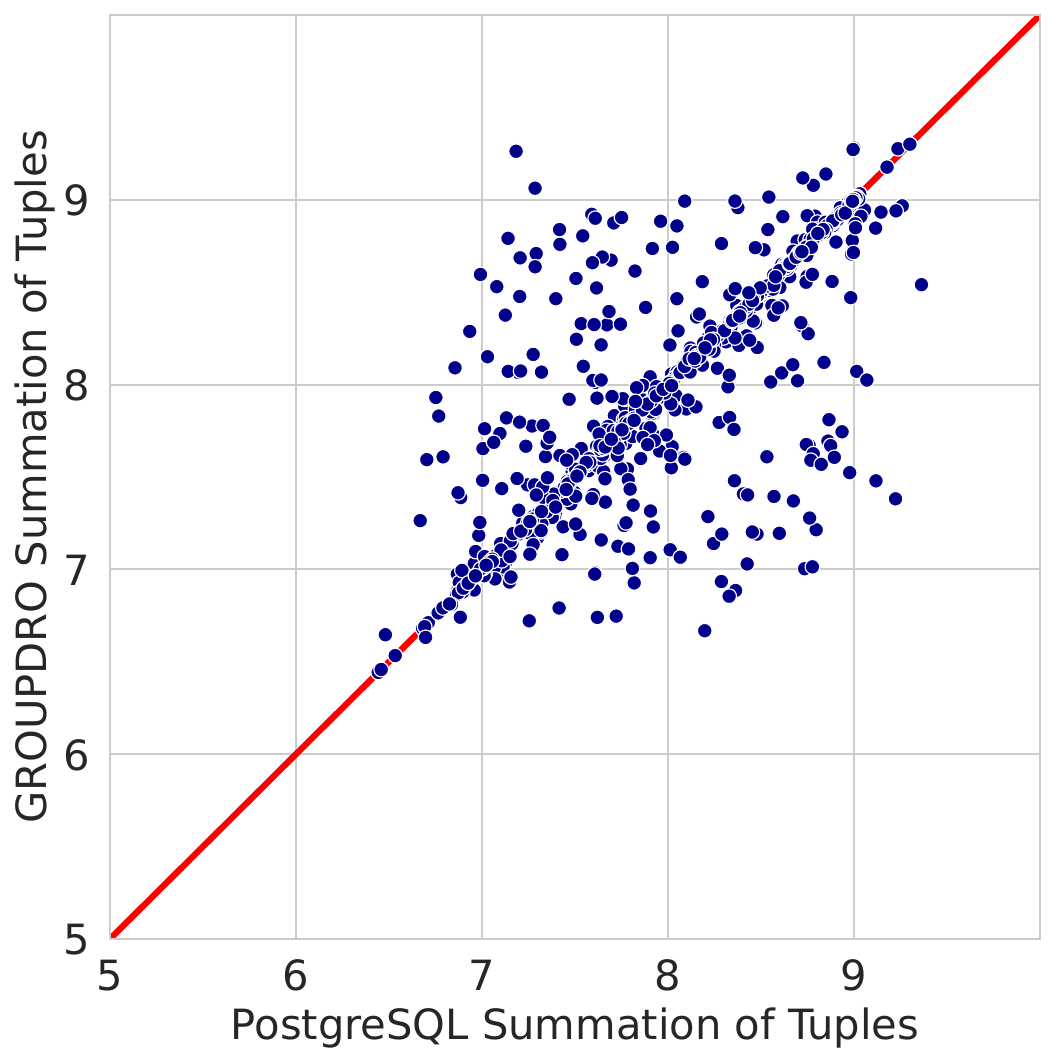}
			\label{fig:tuple_groupdro_pg}
		} 
          \subfigure[mixup] {
				\includegraphics[width=0.28\columnwidth]{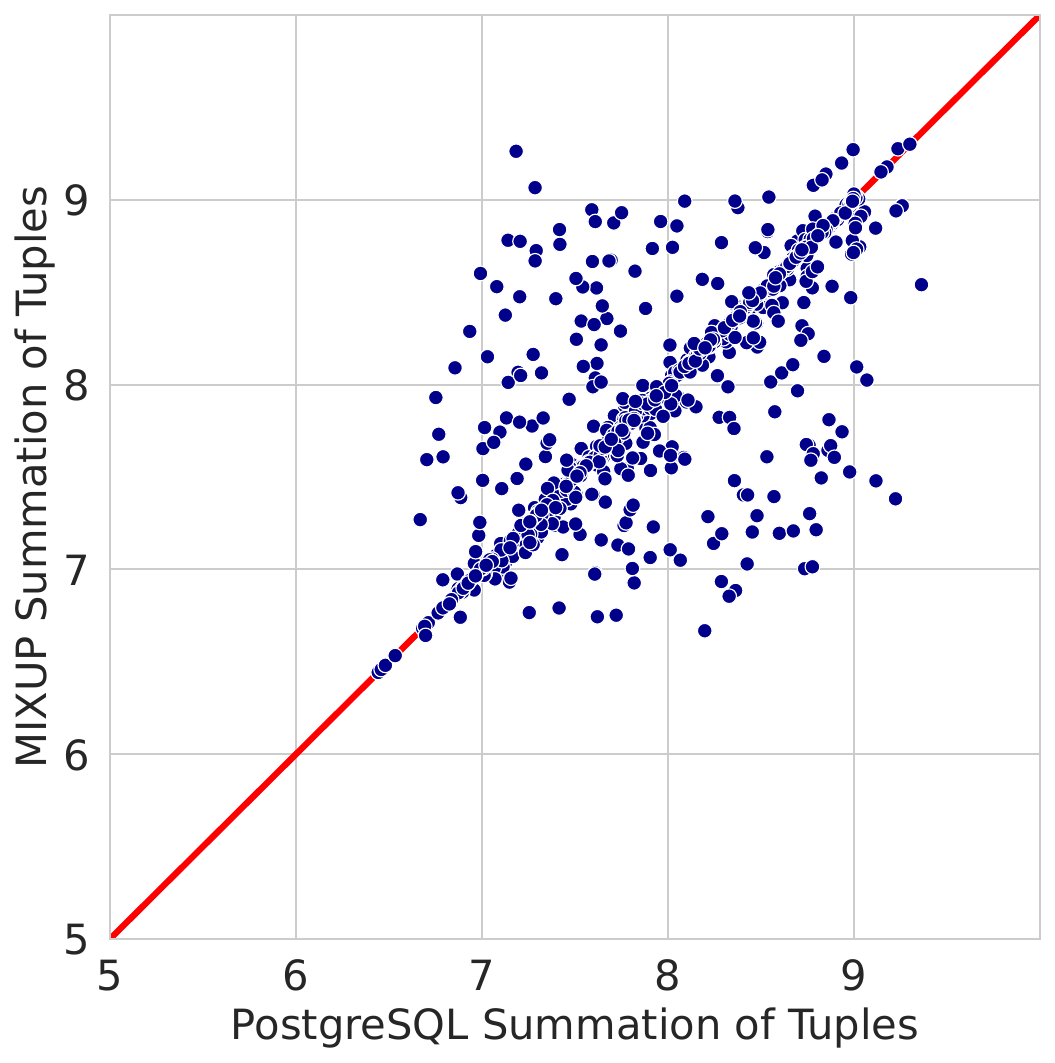}
			\label{fig:tuple_mixup_pg}
		} 
          \subfigure[mask] {
				\includegraphics[width=0.28\columnwidth]{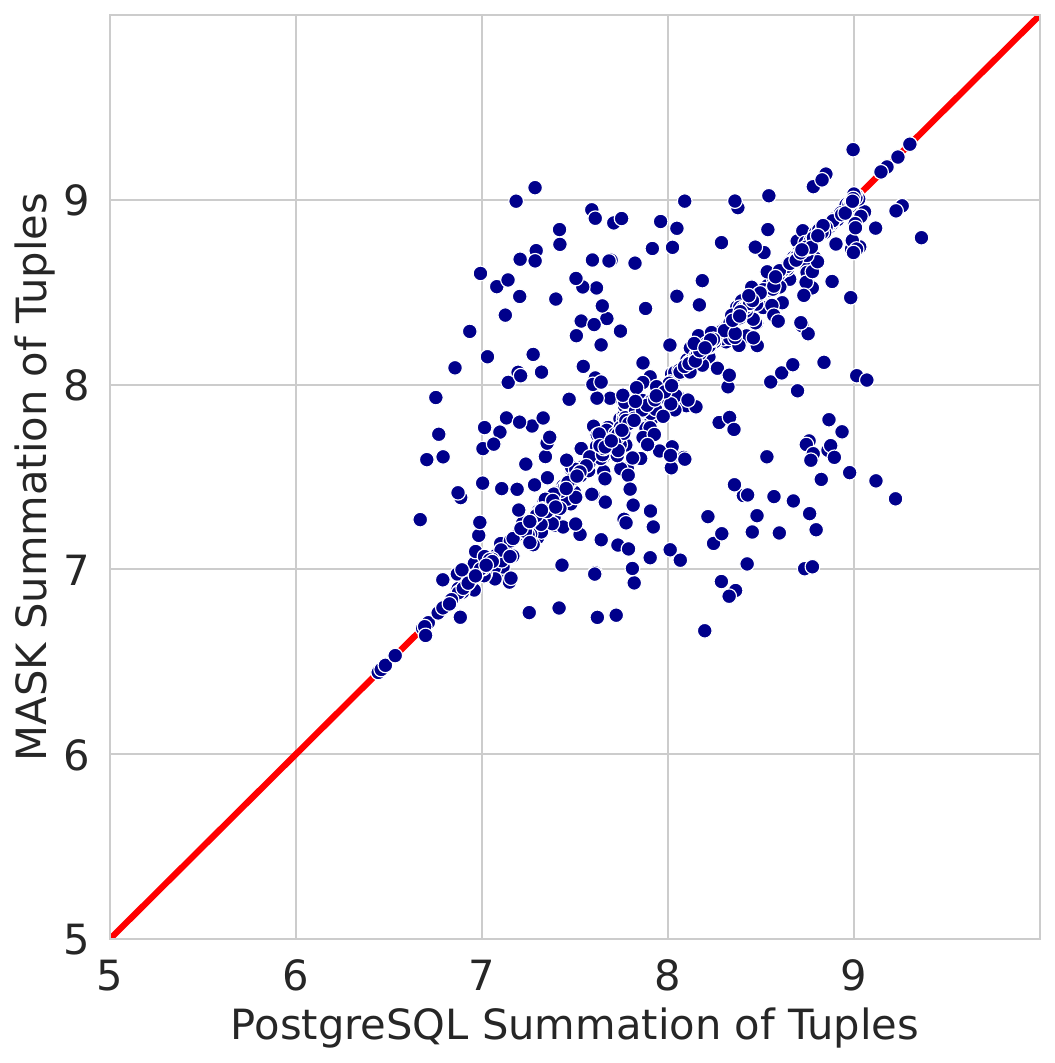}
			\label{fig:tuple_mask_pg}
		} 
	\end{tabular}
  \caption{tuple comparison}
  \label{fig:main}
\end{figure*}
}

\begin{table}[t]
\small 
\caption{Total Query Execution Time (Seconds) in \PostgreSQL}
\label{tbl:query_time}
\begin{center}
\begin{tabular}{crr|rr|rr}
\toprule
 \multirow{ 2}{*}{Estimator}      & \multicolumn{2}{c}{\imdb}    &   \multicolumn{2}{c}{\dsb} 
 &   \multicolumn{2}{c}{\job}\\ 
        & Time            & Ratio          &   Time               & Ratio    & Time & Ratio            \\ \midrule
ERM & $39,267$ & 0.958 &   762 &   0.669 &  $9,887$ & 0.995          \\
Deep CORAL & \textbf{38,690} & \textbf{0.944} & \textbf{722} & \textbf{0.634} & $9,586$ & 0.964  \\
DANN   & $40,094$ & 0.978  & 756 & 0.663 & $9,679$ & 0.974     \\
Group DRO  & $38,955$ & 0.950  & \cellcolor{LightYellow} 726 &  \cellcolor{LightYellow}  0.638   & $9,525$ & 0.958    \\
OrderEmb & \cellcolor{LightYellow} $38,774$ & \cellcolor{LightYellow} 0.946 & 741 & 0.651& $9,655$ & 0.971     \\
Query Mixup & $39,308$ & 0.959 &838 & 0.736  & \cellcolor{LightYellow} 9,490 & \cellcolor{LightYellow} 0.955    \\
Query Masking & $38,910$ & 0.949 & 733 & 0.644 & \textbf{9,479} & \textbf{0.954} \\ \midrule
\PostgreSQL Built-in & $40,968$ & 1.000 & 1,139 & 1.000 & $9,939$ & 1.000    \\ 
True Cardinality & 37,569        &  0.917    & 731 & 0.642     &  9,428            & 0.948           \\\bottomrule
\end{tabular}
\end{center}
\end{table}

%
%
%

\label{sec:exp:postgres}